\documentclass[letterpaper,twocolumn,10pt]{IEEEtran}

\title{Safety vs. Security: Attacking Avionic Systems with Humans in the Loop}

\usepackage{siunitx}
\usepackage{subfig}

\usepackage{textcomp}
\usepackage{eurosym}

\usepackage{multirow}
\usepackage{booktabs}
\usepackage{color}
\usepackage{colortbl}

\newcommand{\greyrow}{\rowcolor[gray]{0.85}}

\usepackage{array}

\usepackage{hyperref}

\usepackage{graphicx}
\usepackage{authblk}

\author[1]{Matthew Smith}
\author[1,2]{Martin Strohmeier}
\author[ ]{Jon Harman}
\author[2]{Vincent Lenders}
\author[1]{Ivan Martinovic}

\affil[1]{Department of Computer Science, University of Oxford}
\affil[ ]{\textit{first.last@cs.ox.ac.uk}}
\affil[2]{armasuisse}
\affil[ ]{\textit{first.last@armasuisse.ch}}

\begin{document}
\maketitle 

\begin{abstract}
Many wireless communications systems found in aircraft lack standard security mechanisms, leaving them fundamentally vulnerable to attack. With affordable software-defined radios available, a novel threat has emerged, allowing a wide range of attackers to easily interfere with wireless avionic systems. Whilst these vulnerabilities are known, concrete attacks that exploit them are still novel and not yet well understood. This is true in particular with regards to their kinetic impact on the handling of the attacked aircraft and consequently its safety.

To investigate this, we invited 30 Airbus A320 type-rated pilots to fly simulator scenarios in which they were subjected to attacks on their avionics. We implement and analyse novel wireless attacks on three safety-related systems: Traffic Collision Avoidance System (TCAS), Ground Proximity Warning System (GPWS) and the Instrument Landing System (ILS). 

We found that all three analysed attack scenarios created significant control impact and cost of disruption through turnarounds, avoidance manoeuvres, and diversions. They further increased workload, distrust in the affected system, and in 38\% of cases caused the attacked safety system to be switched off entirely. All pilots felt the scenarios were useful, with 93.3\% feeling that simulator training for wireless attacks could be valuable.
\end{abstract}

\section{Introduction}
\label{sec:intro}

Over the past few decades, flying has become ever safer, culminating in the year 2017, where not a single death was recorded for commercial passenger air travel \cite{Beeb2018}. As a whole, the aviation industry and regulators have achieved this with a meticulous focus on safety, for example regarding the testing, maintenance and certification requirements of an aircraft.

Two important aspects of this all-permeating safety mindset can be seen both in the continuous training of pilots in flight simulator scenarios, preparing them for safety-critical situations, and also in the numerous wireless technologies on board an aircraft, which are meant to increase situational awareness for pilots and air traffic control (ATC).

If these avionic systems malfunction or are not used as intended, consequences can be serious. Examples include an inoperative transponder on board a Delta Airlines aircraft in March 2011, which remained undetected for ten minutes~\cite{Eurocontrol2014transponder}. During this time it flew in close proximity to three aircraft---a working transponder would have helped avoid this situation. In extreme cases, equipment malfunction can cause loss of life. In 2006, two aircraft collided in Brazil partly due to a failing transponder not providing collision avoidance messages~\cite{AAIPC2006}. 

Similar to many industries with safety-critical components, aviation is currently working on securing their infrastructure against the new threat of cyber attacks. In this process, all wireless technologies have come under scrutiny, as they almost in their entirety lack fundamental security mechanisms \cite{Strohmeier2017}. A subset has been shown to be exploitable under laboratory conditions using widely accessible software-defined radios (SDRs) and software tools (e.g., \cite{Costin2012,Schafer2013,Smith2017}).

Since publication of these proof-of-concept demonstrations, discussion has been ongoing within the aviation sector about the severity and reality of the threat of wireless attacks on safety systems under actual flight conditions. Recent research from the U.S. Department of Homeland Security indicating remote compromise of a Boeing 757 aircraft was dismissed by the manufacturer, who claimed confidence in the security of its aircraft~\cite{Cox2018}. Several surveys on the perspectives of pilots and other aviation professionals highlight that opinion is split. Some believe attackers could succeed in creating `unsafe flight conditions'---the prominent view, however, is that such attacks are mitigated already through aviation's extensive safety systems and culture~\cite{AirLinePilotsAssociationIntl2017,Strohmeier2018}. 

Unfortunately, security research into avionics \cite{Costin2012,Schafer2013,Smith2017,Strohmeier2014a} has already shown that the threat is not fundamentally addressed by safety-oriented design, which deals with random mechanical, electronic, or human failure, rather than deliberate and targeted attempts to subvert the system. Similar to conventional security threats by passengers or pilots, attackers can negatively influence the safety of an aircraft, if they are able to replicate  failures of the wireless avionics systems. 

However, taking a standard security assessment approach to this faces a number of challenges. First, flight hardware is extremely expensive and difficult to use in isolation, making the construction of a real-world test bed prohibitive for independent research. Even more importantly, however, the flight crew have the ultimate authority over how an aircraft is flown, so their response to attacks can create a broad range of impacts from amplifying the effects to mitigating them entirely. Hence, examining wireless attacks with the pilots in the loop is a necessary requirement to gauge their true impact, i.e., the effect on handling and safety of the aircraft.

To quantify this impact, we implement three novel wireless interference attacks in a human in the loop environment. Our work recruits 30 professional airline pilots to fly scenarios in a flight simulator during which they are subject to realistic cyber attacks. The attacks are based on an analysis of theoretical vulnerabilities and real-world interference incidents in three heavily used safety-related systems: the Instrument Landing System (ILS), Traffic Collision Avoidance System (TCAS) and the Ground Proximity Warning System (GPWS).

\paragraph*{Contributions} This paper identifies which systems should be the focus of the security effort as it is the first study to analyse the impact of novel remote wireless attacks with pilots in the loop. Our contributions are as follows:
\begin{itemize}
\setlength\itemsep{0.1em}
\item We describe concrete wireless attacks on avionics, advancing the state of the art for three systems: collision avoidance, instrument landing and ground proximity.
\item We implement these attacks in a flight simulator and run experiments with 30 Airbus A320 pilots in the loop to test their true effect within aviation's safety culture.
\item We analyse in-simulator and interview debrief results from the experiments, quantifying the attacks' kinetic impact on handling and safety of the aircraft as well as potential mitigations and countermeasures.
\end{itemize}

We begin with background in Sec.~\ref{sec:relwork}, before outlining our threat model in Sec.~\ref{sec:threat}. We discuss systems and attacks in Sec.~\ref{sec:systems}, then cover our experimental method in Sec.~\ref{sec:method}. Our results are presented in Sec.~\ref{sec:results}, followed by discussion in Sec.~\ref{sec:discuss}. We provide mitigations and recommendations in Sec.~\ref{sec:mitigate} and conclude in Sec.~\ref{sec:conclude}.

\section{Background}
\label{sec:relwork}
Whilst cyber security in aviation is a more recent concern, investigation in to the effectiveness of flight simulators for training is more developed. In this section we consider the background for both of these areas.

\subsection{Cyber Security in Aviation}
Increasing awareness of cyber threats in aviation has spurred early-stage research into attacks and countermeasures. An early analysis of surveillance system vulnerability generated more widespread attention~\cite{Costin2012}. At the threat modelling level, several works assess feasible types of attack. In~\cite{Johnson2013}, the highlighted threats are spoofing, exploiting, denial of service and counterfeiting. Our study focusses on the spoofing and denial of service attacks. In~\cite{Patel2012}, specific threats are enumerated, including possible consequences of attacks on collision avoidance systems. We directly assess some of these effects.

Furthermore, technical research into the security of secondary surveillance radar (SSR) systems has assessed the constraints on an attacker aiming to inject, modify or delete SSR messages~\cite{Schafer2013}, and provided a thorough assessment of the potential security solutions available~\cite{Strohmeier2014a}. 

Awareness about cyber attacks varies, as demonstrated in~\cite{Strohmeier2017}. The authors survey aviation professionals on their perceptions on the security of a range of different avionic systems. Whilst there is awareness that the systems are not inherently secure, there does not appear to be significant concern that attacks could affect operational capability.

\subsection{Simulator Training}

Time spent in the simulator is a vital part of professional pilot training. A body of research analyzes the configuration of simulator scenarios such that they transfer most easily to flying the real aircraft. Early research indicated that it provides notable benefit over aircraft-only training~\cite{Hays1992}. However, it is not a given that high-fidelity simulation transfer skills well, and the literature suggests that well-designed scenarios are vital in equipping pilots effectively~\cite{Salas1998,Dahlstrom2009}.

One of the key factors in cyber attacks is that there may be no forewarning, leading to surprise and loss of capacity. In~\cite{Kochan2004}, a survey of aviation incident reports highlights that `normal' events can be surprising to pilots when they occur out of context, i.e. alerts when the conditions do not warrant it. The authors in~\cite{Landman2017,Casner2013} consider this with respect to stall recovery manoeuvres, a regularly tested skill for pilots. Both papers find that pilots struggled to follow even well-known procedures when the stall occurred in unexpected conditions. 

Addressing this, the authors of~\cite{Landman2018} argue that unpredictability and variability in simulator training improves performance when encountering surprise scenarios. While their work uses failure scenarios instead of malicious interference, the arguments remain valid.

\subsection{Simulating Cyber Attacks}

Some work addressing simulation for cyber security has begun to emerge. In~\cite{Gontar2018} the authors conduct a human factors focussed study to assess how pilots respond to an attack on ground-based navigation systems. They find that pilots under attack lose some monitoring capacity, and that warnings can help mitigate this. The authors of~\cite{Buch2019} (and the extended~\cite{EASA2018}) conduct a more avionics-focussed set of attacks, looking at six variants of navigation and flight management system threats. Multiple attacks inserted over the course of one flight with the intention being to assess if pilots notice the attacks. They found that most attacks were identified during flight, however some happened without detection.

Our work differs in that we focus specifically on systems that are either entirely or partly safety-critical in their usage, with the attacker instead aiming to disrupt. We also choose to explore a different set of systems and cover the principles of these attacks in technical detail.

\section{Threat Model}
\label{sec:threat}
We presume a moderately resourced attacker, with a budget in the region of \$10-15,000 to buy commercially off-the-shelf antennae, amplifiers, and SDRs. This would enable them to transmit at sufficient power to communicate with airborne aircraft. We also presume they have the capability to develop software, or use existing open-source tools, to interfere with aircraft systems. Our attacker can deploy their systems remotely or create a mobile platform from which to do this.

\paragraph{Threat Actors.} We consider three threat actors: activists, terrorists and nation states. Activists intend to cause disruption to raise the profile of their cause, usually with low resource but high levels of personnel. On the other hand, terrorists aim to disrupt or destroy with the intent of creating a chilling effect or fear. They can be moderately resourced and are unlikely to care about collateral damage. Most extreme is the nation state who primarily intend to disrupt in order to paralyse infrastructure. They are well resourced and are likely to be concerned about attribution and collateral.

\paragraph{Attack Aims.} We focus on attempts to cause disruption, rather than destructive impact. This is due to our work looking to the fundamentals of the attacks, in which disruption is likely to be the first effect, though these likely have destructive variants. This can include diversions to alternative airports, excessive movement away from planned routes or \textit{go-arounds}, i.e. a missed approach to land followed by a second attempt. This work performs a feasibility analysis on the effects of these attacks, which is indicative of the impact they would have under a stronger threat model. As such, we believe that based on our results, future work could focus on this aspect. We discuss this further in Sec.~\ref{sec:discuss}.

Furthermore, we are careful to ensure the experiment is fair. In scenarios where the aircraft is put at risk of crashing it would be unrealistic to assess pilot response outside of their normal environment. For example, we could not accurately assess response times if controls are slightly different to a full simulator or real aircraft. We cover our experimental setup and its limitations in Sec.~\ref{sec:method}.

\section{Systems and Attacks}
\label{sec:systems}
We now describe the systems and theory of attacks used in the experiment, including attacker capabilities and expected crew responses.

\subsection{Ground Proximity Warning System}

A fundamental part of an aircraft's `safety net', the Ground Proximity Warning System (GPWS) provides early warning of the aircraft becoming too close to terrain~\cite{Breen2015}.

\subsubsection{System Description}

Two versions of this system exist--- the original GPWS, and the newer Enhanced GPWS (EGPWS) which incorporates GPS and a terrain database. The subsystem used in this study is the same in both. Taking a range of sensor inputs, GPWS provides alarms of situations leading to collision with terrain~\cite{UKCAA1976}. It has a range of alert modes; we focus on excessive closure on terrain, or Mode 2~\cite{Breen2015}. 

\begin{figure}[t]
\centering
\includegraphics[width=\columnwidth]{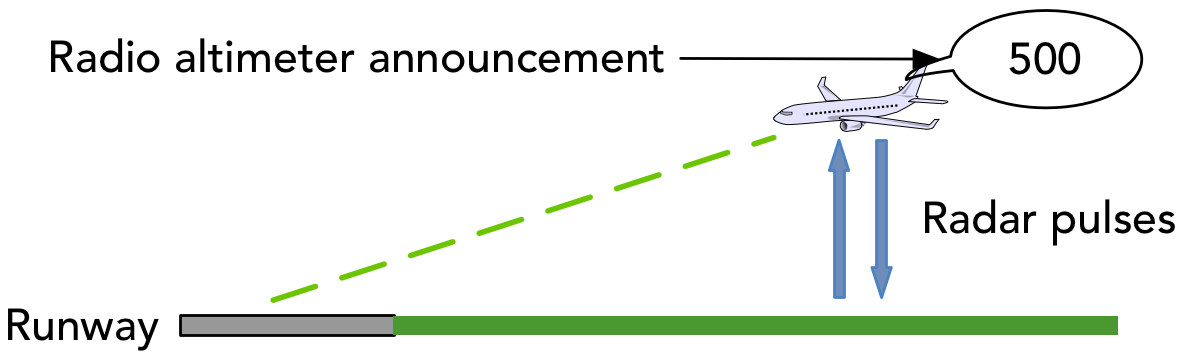}
\caption{Normal system operation with radio altimeter determining height above ground.}
\label{fig:gpws}

\end{figure}

\begin{figure}[t]
\includegraphics[width=\columnwidth]{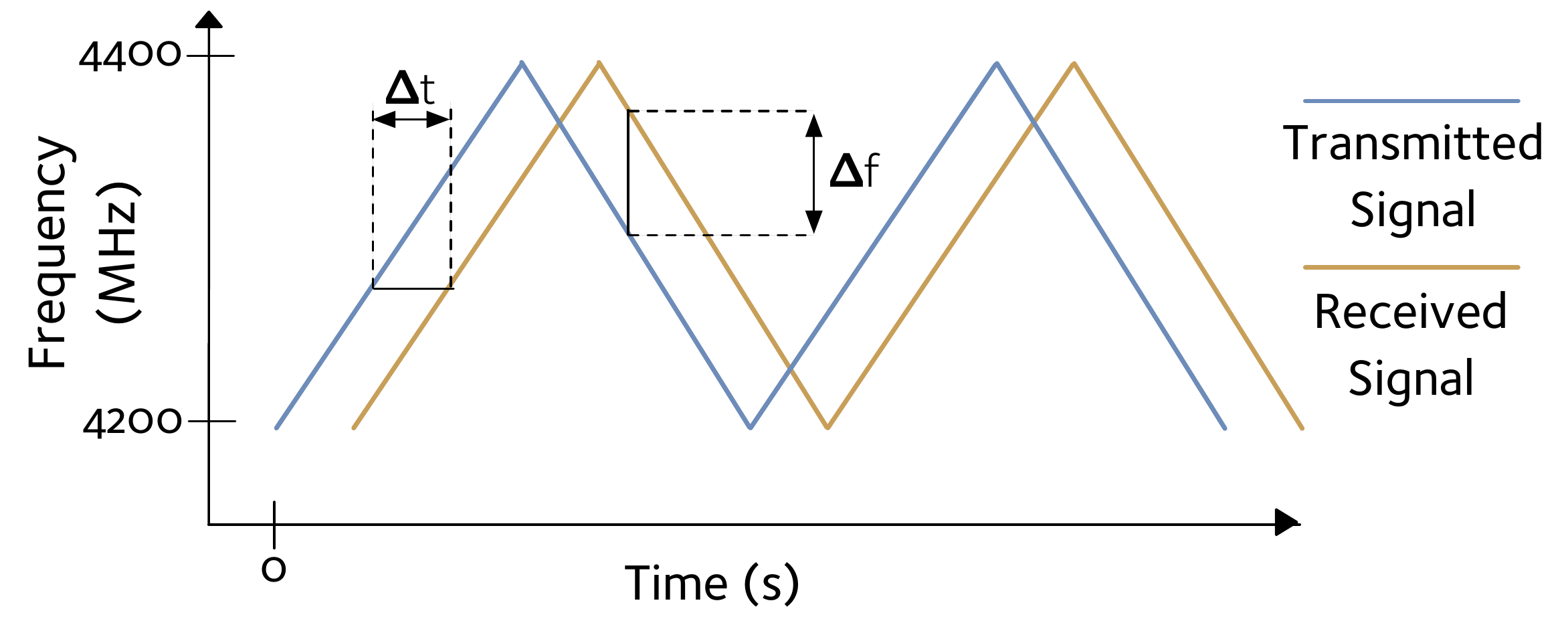}
\caption{Frequency-modulated continuous wave (FMCW) radar operation at a static height, for both the transmitted signal and received, reflected signal.}
\label{fig:sim-fmcw}

\end{figure}

Mode 2 GPWS uses a radio altimeter to determine the rate of closure on nearby terrain; we provide a representation of this on approach in Fig.~\ref{fig:gpws}. A radio altimeter is a Frequency-Modulated Continuous Wave (FMCW) radar, transmitting pulses on frequency sweeps between 4200 and 4400~MHz. It uses the frequency shift and round trip time for the received signal to calculate the height above terrain, also referred to as \textit{above ground level} (AGL). Its operation is illustrated in Fig.~\ref{fig:sim-fmcw}, where $\Delta t$ is the round trip time, and $\Delta f$ is the frequency shift. 

Mode 2 has two sub-modes, A and B: A is primarily active during climb and cruise whilst B is used during approach and landing. Excessive terrain closure will be met with audio alerts, the most serious of which is \textit{`Terrain Terrain, Pull Up.'} GPWS is considered a high priority alert for crew~\cite{Skybrary2018pullup}.

\subsubsection{Attack Description}
Our attacker aims to create a spurious alert to negatively impact situational awareness and  cause an unwarranted go-around. As a result, aircraft will then have perform a second approach or divert to a different airport. During this time, the aircraft will be using extra fuel, incurring delay, as well as adding workload for the pilots. By transmitting false radar pulses on final approach, the attacker causes the GPWS to believe that the terrain closure rate is significantly higher than reality. This will trigger a \textit{`Terrain Terrain, Pull Up'} whilst the aircraft is close to the ground yet within a `safe' range.

Two methods could enable this attack. A targeted attack aims to replicate the rapid closing of ground by transmitting a ramp of frequencies between 4200\,MHz and 4400\,MHz. The gradient of this ramp is crafted to incrementally reduce the round trip time per frequency shift for the signal, creating the illusion of the ground approaching rapidly.

This requires some prediction of the signal phase from the radio altimeter, as well as knowledge of the sweep frequency---however this is standardized. Since Mode 2 alerts are based on the rate of descent, the attacker can at least calculate the target change in  round trip time (RTT) to trigger an alarm. For example, descending at 3000\,ft/min (\textasciitilde 15.4\,m/s) at 500\,ft AGL (\textasciitilde 152.4\,m) will trigger an alarm according to standard (Fig. A2b in~\cite{UKCAA1976}). Using a simple model of the aircraft moving a negligible amount during a pulse, we use the difference in RTT over the course of one second (i.e. the aircraft at 152.4\,m AGL, then one second later having descended 15.4\,m). Eq.~\ref{eq:timedelta} then gives us the required change in is RTT, in which $t_{rtt}$ is RTT, $h$ is height above ground and $c$ is the speed of light. This indicates a small jump in frequency per round trip is needed.

\begin{equation}
\Delta t_{rtt} = \frac{2(h_{1}-h_{2})}{c}= \frac{2(152.4-137.0)}{c}\approx 1.03 \time 10^{-7} \text{s/m}
\label{eq:timedelta}
\end{equation}

Another approach is to flood the aircraft with many signals within the radar frequency range, either at a specific frequency or randomly. The behaviour of the radio altimeter in this scenario is unknown as it would be receiving effectively unpredictable pulses, so we do not consider it further.

\paragraph*{Expected Response}
Whilst the response will depend on the aircraft and airline, there are common principles~\cite{Skybrary2018pullup}. In most conditions we expect a terrain avoidance manoeuvre, i.e. a steep climb to a safe altitude. In our scenario, this will lead to a missed approach. However, below 1000\,ft above aerodrome level (AAL), with full certainty of position, crew can choose to not follow this. Due to the surprise element, we expect the typical response to be a missed approach. On following approaches we expect participants to have identified unexpected behaviour and disregard the warnings.

\subsubsection{Requirements \& Feasibility}
For this, an attacker will need a number of directional antenna underneath the approach path to transmit to the radio altimeter. These will be fed by SDRs and software capable of transmitting correctly modulated pulses; such equipment would be in the low \$1000s. Although an attacker could operate such a system remotely, the hardware would need to be located near to the runway. 

The ability to deploy depends on the airfield security and perimeter size. Whilst an attacker could cause maximum confusion with a late alarm, this attack can take place from afar. Using a representative 130\,knots landing speed descending at a standard 700\,ft/min, an attack at 500\,ft AGL would occur around 1.7\,miles from the start of the runway.\footnote{This is calculated based on the aircraft taking $\scriptstyle\mathtt{\sim}$42\,s to reach the touchdown zone.}

With regards to system resilience, GPWS radio altimeter is known to suffer interference from outside sources. In~\cite{Hovav2011}, instances of unexplained GPWS alarms on approach to an airport were explained by emissions from a nearby military radar. Outside interference on radio altimeter frequencies has been occuring for some time, with discussions ongoing due to the importance of the instrument~\cite{ICAO2018}. As such, a targeted attacker could have a greater effect.

\subsubsection{Simulator Implementation} 
We simulate the attack by triggering the GPWS `\textit{Terrain, Terrain, Pull Up}' alarm starting 500\,ft AGL on approach to Runway 33 at Birmingham, increasing by 250\,ft for each subsequent attack. With this we are emulating the ability of an attacker to add some unpredictability to the attack. One of the limitations of this approach is that the point at which the attack actually triggers can vary between 450\,ft and 500\,ft AGL, and the radio altimeter visual does not show an `under attack' change for the time under attack.

\subsection{Traffic Collision Avoidance System}
\begin{figure}
\centering
\includegraphics[width=\columnwidth]{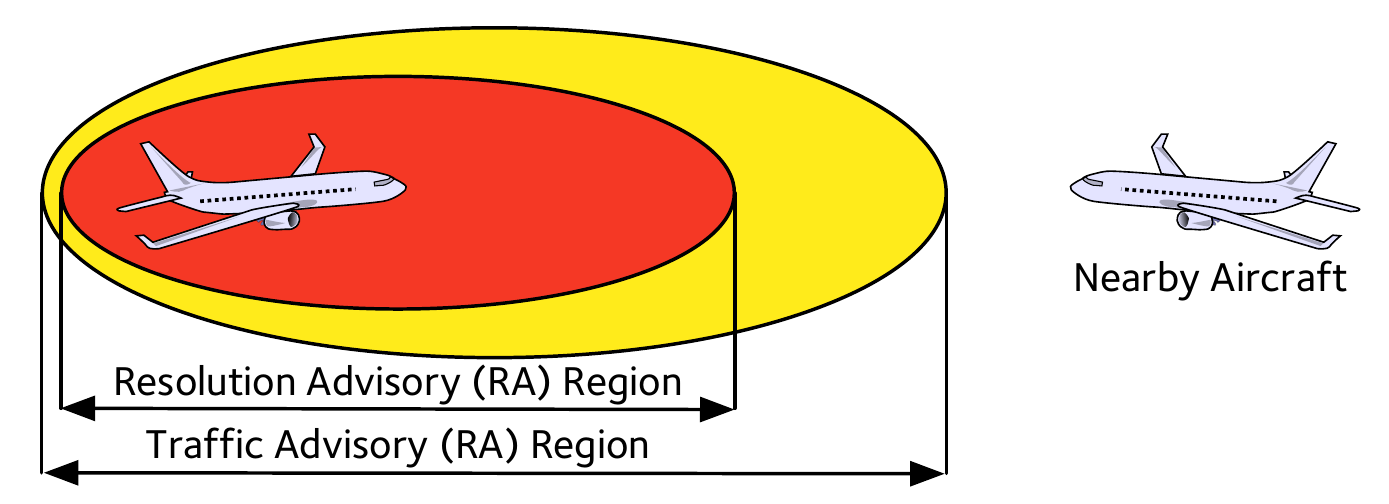}
\caption[Representation of TCAS Traffic (TA) and Resolution Advisory (RA) zones.]{Representation of TCAS Traffic (TA) and Resolution Advisory (RA) zones.}
\label{fig:sim-TCAS-rings}

\end{figure}

Although ATC manage airspace with high precision, aircraft can still end up closer than is safe. This is called a \textit{loss of separation}, and in the worst case, can result in a mid-air collision. One such example occured in March 2011, where a Delta aircraft took off with an inactive transponder, becoming too close to three other aircraft before resolving the issue~\cite{Eurocontrol2014transponder}. Traffic Collision Avoidance System (TCAS) provides a technical means to avoid this, and has been mandated on aircraft with more than 30 seats since 1993~\cite{FAA2011a, Henely2015}.

\subsubsection{System Description}
\begin{figure}[t]
\centering
\subfloat[Protocol diagram for TCAS interrogation using the Mode S data link, where nearby aircraft respond with information on their position.]{
\includegraphics[width=\columnwidth]{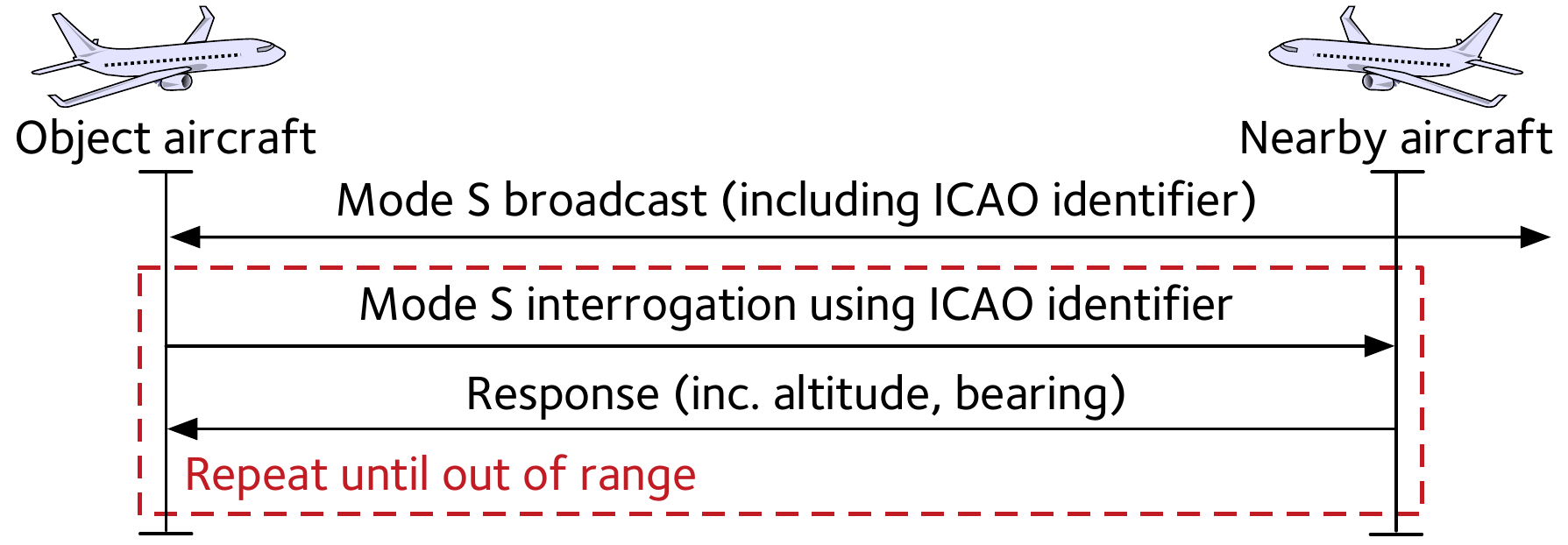}
\label{fig:sim-tcas-proto}
}
\\
\subfloat[Protocol diagram of TCAS all-call interrogation using Mode C, and response from nearby aircraft with altitude if available. Range and bearing are calculated from response.]{
\includegraphics[width=\columnwidth]{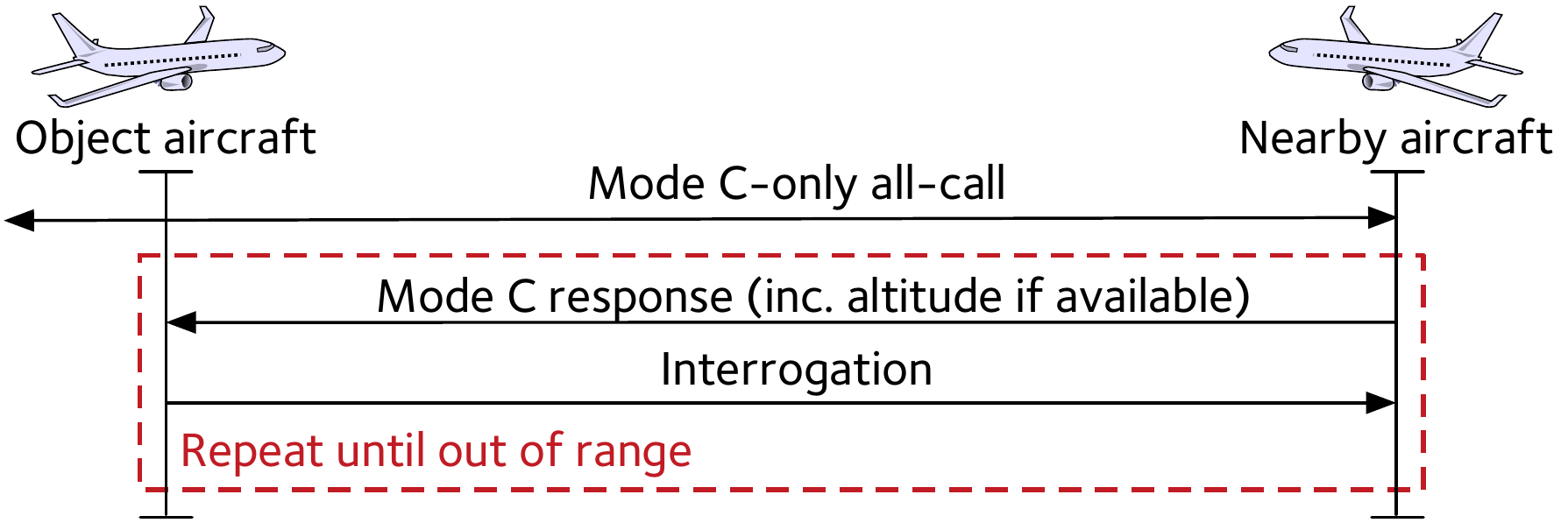}
\label{fig:basic-tcas}
}
\caption[Representation of TCAS interrogation protocols of nearby aircraft using Mode C and S transponders.]{Representation of TCAS interrogation protocols of nearby aircraft using Mode C and S transponders.}

\end{figure}

TCAS makes use of the Mode C or Mode S transponders fitted to an object aircraft to interrogate nearby aircraft~\cite{FAA2011a}. Establishing nearby aircraft with Mode S requires the object aircraft to listen for IDs in Mode S `squitters', which are messages in response to ground-based Secondary Surveillance Radar (SSR) interrogations. The object aircraft can then interrogate these IDs to calculate whether nearby aircraft will become too close~\cite{FAA2011TCASModeS}. An abstracted protocol diagram for Mode S can be seen in Fig.~\ref{fig:sim-tcas-proto}.

Mode C operates differently, shown in Fig.~\ref{fig:basic-tcas}. The object aircraft issues Mode C-only interrogations called \textit{all-calls}, causing all nearby aircraft with Mode C transponders to respond once per second with their altitude. Since Mode C does not carry the same data fields as Mode S, the object aircraft estimates range and bearing~\cite{FAA2011TCASModeC}. To limit interference it uses a \textit{whisper-shout} transmission mechanism, gradually increasing power and suppressing aircraft who have already responded.

Based on lateral and vertical proximity to nearby aircraft, visual representation and alerts are given to crew. These come in two steps as represented in Fig.~\ref{fig:sim-TCAS-rings}. First is a \textit{traffic advisory} (TA), in which the traffic is typically displayed to the pilot as amber and an aural alert of `traffic' is given. If the nearby aircraft becomes closer to the aircraft, a \textit{resolution advisory} (RA) is given. An RA will contain specific instructions for the flight crew, i.e., to climb or descend at a given rate, or hold vertical speed. These instructions are decided between the two aircraft automatically to deconflict the situation. RA instructions must be followed within seconds.

In the cockpit, crew have control over the alerting level; they can select \textit{Standby}, \textit{TA-ONLY}, or \textit{TA/RA}. For most of a flight, TCAS will be set to TA/RA in which full alerting is provided. TA-ONLY does not issue RAs, whereas Standby performs no TCAS interrogations or conflict resolution~\cite{FAA2011TCASTau}.

\subsubsection{Attack Description}

In this scenario our attacker aims to cause crew responses to TCAS by triggering TAs and RAs despite no aircraft being nearby. This is intended to burn unnecessary fuel, break from air traffic control clearances and cause knock-on alerts for other aircraft, possibly resulting in diversions or switching TCAS off. This is due to Mode C and S transmissions being sent in the clear with no authentication. Sch\"{a}fer et. al. use a similar concept on a system which uses Mode S, the concepts of which we translate to TCAS~\cite{Schafer2013}.

An attacker generates Mode C/S TCAS responses for a false intruder aircraft gradually approach the object aircraft from a distance, with the closest point of approach being sufficient for alarm. We will refer to the aircraft under attack as \textit{target} and the injected aircraft as \textit{false}. 

We firstly presume that we can establish the altitude, heading and speed of the target aircraft from broadcast surveillance messages~\cite{Strohmeier2014a}. Injection then depends on whether they use Mode S or Mode C:
\begin{itemize}
\setlength\itemsep{0.1em}
\item \textbf{Mode S}: the attacker transmits a false aircraft squitter message. When the target aircraft then interrogates, the attacker transmits Mode S responses as if the false aircraft were travelling on a collision course with the target. 
\item \textbf{Mode C}: the attacker responds to an all-call and following interrogations for the false aircraft. Whisper-shout may cause interrogations to be too low power to be received by the attacker, in which case they would need to approximate a response. However, this would be stochastic as interrogation rates are standardized.
\end{itemize}
The attacker can choose whether to cause the target to climb or descend by injecting an aircraft below or above the object aircraft respectively.

A different approach would be to flood the frequencies with Mode C/S responses from a false aircraft. However, the 1090\,MHz link is liable to overcrowding and message loss, meaning that flooding could simply jam the frequency~\cite{Schafer2014}. 

\paragraph*{Expected Response}
Since following an RA is compulsory, we expect that most pilots will comply with at least the first instance thus following the instructed manouvre~\cite{Skybrary2018tcas}. From there we expect some participants to doubt RAs and eventually turn the alert level down from TA/RA to TA-Only or Standby. On average, we expect participants to follow the first 3-4 RAs before reducing the alert level or switching the system off.

\subsubsection{Requirements \& Feasibility}

Transmission by the attacker would require an off-the-shelf amplifier and antenna capable of directional transmission, with a high powered setup costing \$\num{15000}. The attacker would also need software to undertake the attack. A transceiver is needed to both receive interrogations to establish the target aircraft behaviour, as well as transmitting false aircraft messages. Management software could then create the attack by establishing the position of the aircraft and messages needed to cause an alarm.

With regards to software, some related tools exist to decode Mode S signals such as \texttt{gr-airmodes} or \texttt{dump1090}~\cite{Sanfilippo2017,OpenskyNetwork2018}. Encoders are rarer with one such example being ADSB-Out~\cite{Yusupov2017,Young2018}. However, this is not general purpose. An attacker would either need to build such a tool or rely on the hobbyist community producing one open source.


Analysis of the TCAS II logic suggests that attacks creating situations similar to transponder failure can have a range of effects~\cite{Eurocontrol2014failure}. Most related to this scenario are intermittent Mode C transmissions or duplicated Mode S addresses. For the former, the target aircraft may generate late alerts due to the system treating it as not providing altitude. Of the latter, TCAS will ignore the more distance duplicate address, which could be used to deny situational awareness by an attacker.

Physical location is important in this attack in order to maximize range. By design, Mode C and S messages can be transmitted and received on the ground by surveillance radars. To achieve this, aircraft use a relatively high transmission power of up to 250\,W with groundward directional antennae, meaning that with sufficient power this attack can occur from the ground~\cite{ARINC2015}. Because of aircraft altitude, the potential area in which the attacker can reside is quite large and can be maximised by locating on high ground or near an airport.

One of the main feasibility challenges of this attack is that the aircraft may quickly move out of attack range due to its high speed. A well-resourced attacker might deploy antenna to multiple locations, whereas a more simplistic attacker could instead seek higher ground to maximise range. 

\subsubsection{Simulator Implementation}
Within the simulator, we enact a strong attacker who covers a large geographic area, attempting to trigger 10 alerts over the course of the flight. We varied the angle and speed of approach by the false aircraft. We configured these to be identical for each participant. False aircraft began to be injected when the target aircraft flew above 2000\,ft, after which the first injection began. If the participant chose to turn the TCAS sensitivity to TA-Only, they would still receive TAs but not RAs. 

This attack was undertaken by using an invisible aircraft model which travelled towards the target aircraft. Further work would improve the realism of this, such as more realistic flight patterns to avoid tipping off participants to the attack. 

\subsection{Instrument Landing System}
The Instrument Landing System (ILS) allows precision landings even in poor weather conditions. Since aircraft must follow specific arrival routes into an airport, ILS is an important part of managing pilot workload and is the default approach type for most airports. In extreme cases, ILS allows aircraft to automatically land at sufficiently equipped airfields.

\subsubsection{System Description}

ILS consists of two components: localizer (LOC) and glideslope (GS)~\cite{FAA2012}. A localizer provides lateral guidance and alignment, centered on the runway centerline, whereas the GS provides vertical guidance to a touchdown zone on the runway. Typically the GS will provide a \ang{3} approach path, though this depends on the specific approach and airport~\cite{Saul-Pooley2017gs}. It is supplemented by Distance Measuring Equipment (DME), which provides the direct distance to a beacon without directionality. 

\begin{figure}[t]
\centering
\subfloat[Glideslope under normal operation.]{%
	\includegraphics[width=\columnwidth]{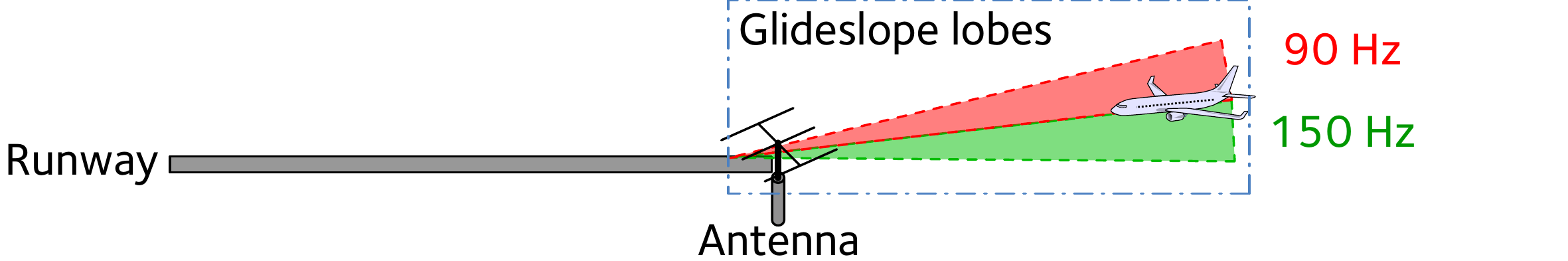}
	\label{fig:glideslope}
}
\\
\subfloat[Glideslope under attack with rogue antenna. Note how the aircraft touchdown zone is now at the far end of the runway. This means that if the glideslope is followed to touchdown, there may not be enough runway to slow down.]{%
	\includegraphics[width=\columnwidth]{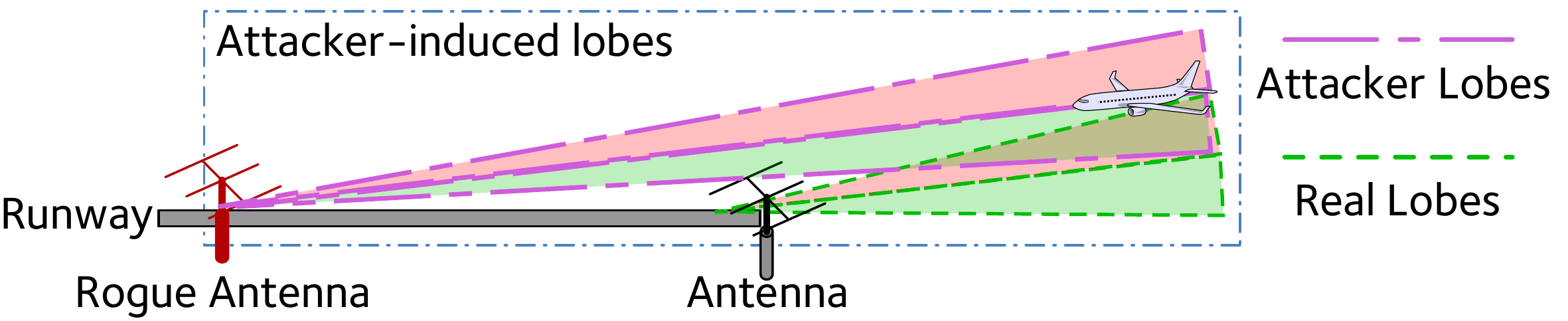}
	\label{fig:glideslope-attack}
}
\caption[Representation of normal and under-attack glideslope operation.]{Representation of normal and under-attack glideslope operation, based on diagrams from~\cite{FAA2012}.}

\end{figure}

Transmission powers of the GS and LOC are 5\,W  and 100\,W respectively~\cite{FAA2012}. On the carrier frequencies for the GS and LOC overlapping 90\,Hz and 150\,Hz lobes provide guidance with the overlap forming the correct approach path. The aircraft will use the relative strength of these lobes to identify where it is with respect to the optimal GS and centerline of the runway. A diagram of a GS can be seen in Fig.~\ref{fig:glideslope}.

GSs and LOCs are monitored for accuracy to at least 10\,nm beyond the runway, as well as being protected from interference to 25\,nm~\cite{Saul-Pooley2017loc,Saul-Pooley2017gs}. It is important to note that `protection from interference' here means avoiding systems using nearby frequencies, rather than malicious interference.

Separately, approach lighting provides an out-of-band check for crew on approach---Precision Approach Path Indicators (PAPIs) are configured to match to the angle of the GS. When an aircraft is on the correct GS, the PAPIs will show two red and two white lights, otherwise more red or white lights are shown as appropriate~\cite{FAA2015}. 

\subsubsection{Attack Description}

Here, the attacker is aiming to cause unnecessary missed approaches as a result of a tampered GS. In turn, this will use additional fuel, introduce delay and potentially force aircraft to divert to a different airport. A secondary aim might be to force crew to use a different, also attacked, approach method.

The attacker replicates the real GS but with the touchdown zone short or long of the legitimate touchdown zone by transmitting a replica signals from aside the runway. Since they will not be able to station themselves on the runway, they will operate outside the airfield perimeter. This somewhat matches the legitimate GS signal which is transmitted aside the runway to avoid aircraft clipping the antennae. 

Crucially the signals would be the same as a real GS, so would not be identifiable by a high rate of descent, as common GS issues can be. The difference induced by the attacker would be subtle. For a typical \ang{3} GS, moving it 1\,km along the runway creates a consistent height difference between the real and false GS of approximately 52\,m, or 172\,ft. This could fall within a margin of error on approach, especially whilst further away from the runway.

\paragraph*{Expected Response}
Since this attack will see the false GS track slightly above the real GS, it is unlikely to be immediately obvious that it is incorrect. We expect most participants to follow the GS until they are below cloud at around 1000\,ft, at which point they will notice a continued slight discrepancy in AGL according to approach charts. They may also notice such a discrepancy using the PAPI, as they will show four white lights. At this point, we expect them to be between 500--1000\,ft AGL and opt for a missed approach and go around.

\subsubsection{Requirements \& Feasibility}
No significant technical barriers exist for this attack. This is possible due to the simplistic nature of the system---whilst it is monitored for integrity as defined in ICAO Annex 10, this is for deviations in the legitimate signal rather than malicious interference~\cite{ICAO2006}. An ILS system will normally shut down or notify ATC if excessive deviation is identified.

An attacker will need an SDR, amplifier and directional antennae to replicate the antennae arrays used for the legitimate GS, costing around \$10,000. Since no open-source tools exist to do this, software would need to be created but this is achievable by moderately resourced attacker as it involves implementing a standardised, static system. Furthermore, the transmission power is readily achievable with consumer amplifiers as a typical GS is below 10\,W. For reference, even the lowest level of licensed UK amateur radio operators can transmit in frequency bands surrounding aviation bands at up to 10\,W~\cite{Ofcom2018}. 

Related work suggests that ILS course deviation attacks are possible using such equipment. Sathaye et al. describes two signal generation approaches which enable ILS signal interference, leading to an attacker being able to adjust the localizer or glideslope as seen in the cockpit~\cite{Sathaye2019}.

Although extensively used and relied upon, ILS signals face challenges due to their relatively simplicity which indicates that they are not entirely robust. The two examples as described in~\cite{FAA2012ILSError} are:
\begin{itemize}
\setlength\itemsep{0.1em}
\item \textbf{False lobes}: lower-lobe GS signals reflecting off the ground and appearing to be a legitimate GS according to cockpit instruments. They are identified by their unduly steep angle of descent, usually $9-12^{\circ}$~\cite{FAA2012ILSError}.
\item \textbf{Interference}: namely reflections off buildings or vehicles, or other aircraft moving through ILS signals and distorting them~\cite{Trautvetter2012}. Pilots are taught to expect this kind of interference~\cite{Miller2012}.
\end{itemize}

Challenges lie in physical location and limitations introduced by monitoring. The attacker will have to locate close to the airport perimeter to have correct signal directionality along the runway. However, since airports are usually accessible by road, this should be possible. 

Due to a lack of public information on monitoring makes it difficult to determine whether transmitting under the legitimate GS will trigger an alarm. Depending on attacker aim, care may be required to avoid interference with the legitimate GS, thus shutting the ILS down. We instead consider the `long' version of the attack, where an attacker moves the touchdown zone to the far end of the runway, attempting go-arounds by leaving the aircraft too high correct the approach and land with enough runway to stop. This does introduce a limitation in that it relies on aircraft intercepting the GS from above, which whilst not preferred is a valid method of intercepting the GS~\cite{Saul-Pooley2017gs}. This would also mean the false GS is transmitted above monitoring.

\subsubsection{Simulator Implementation}

In the simulator, an attacker transmits a false GS at the far end of the runway with an effective shift of 2.05\,km, or 1.27\,miles, creating a difference between the false and true GS of 107\,m, or 352\,ft. Due to the way in which ILS is implemented in the simulator software, we could not replicate also having a `real' GS. To account for this we operated on an assumption that the attacker transmits at a higher power than the real GS in an effort to force capture on to the false GS. The manipulation remains in place regardless of how many approaches are made. We treat the participant aircraft as if it is the first to encounter the attack, with ATC not observing previous aircraft having difficulties.

\section{Experimental Method}
\label{sec:method}

\begin{figure}
\centering
\includegraphics[width=\columnwidth]{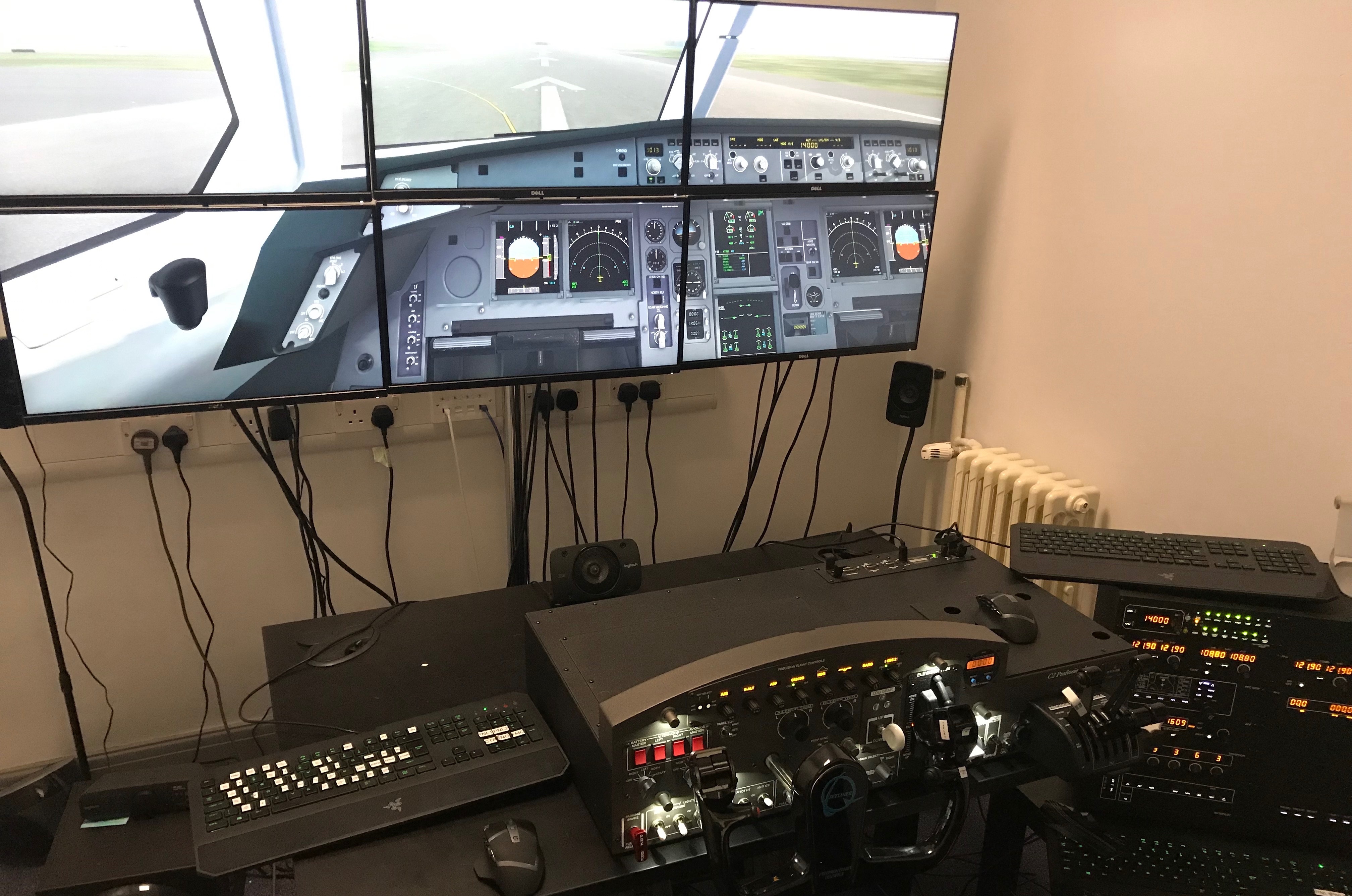}
\caption{Picture of experimental setup.}
\label{fig:setup-pic}

\end{figure}

Since our attacks were specifically designed to examine responses we wanted to allow participants to react in real time. To do this we used a flight simulator, partially recreating a cockpit environment---in this section, we describe the experimental setup used. The work was approved by our local ethics committee with reference number R54139/001.

\begin{figure*}[t]
\centering
\includegraphics[width=\textwidth, trim=17mm 231mm 15mm 17mm, clip]{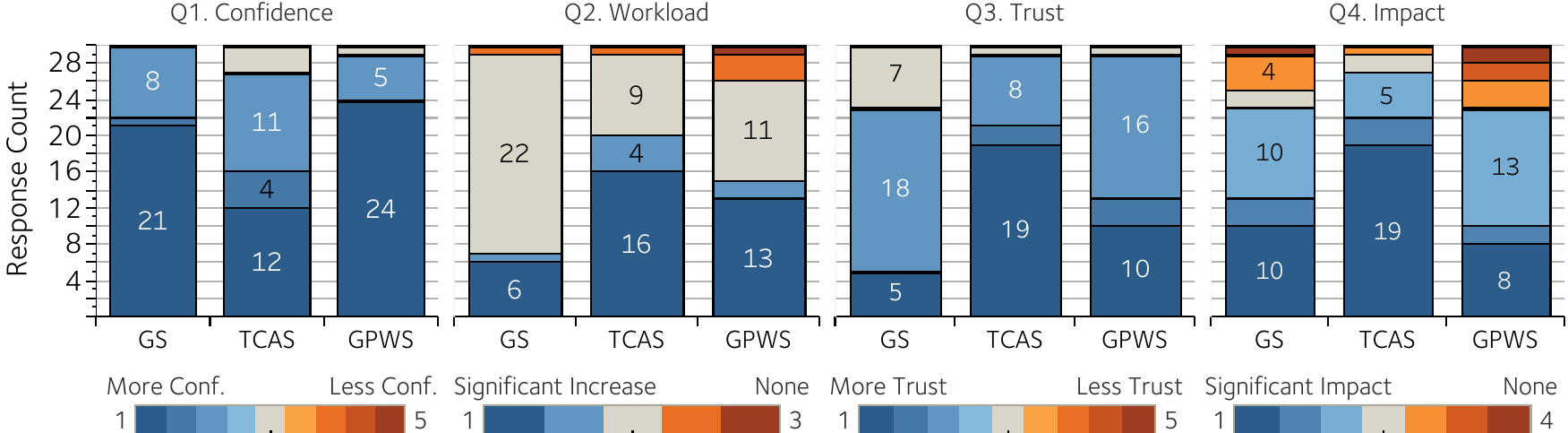}
\caption[Stacked bar charts for simulator participant scale responses.]{Stacked bar charts for participant scale responses on Q1-4. Orange represents the most `negative' responses, i.e. no effect, with blue `positive', i.e. significant effect. Tabular data is provided in Appendix~\ref{app:interview}.}
\label{fig:interview}
\end{figure*}

\subsection{Participants}
We recruited 30 pilots who had current A320 type-rating, or had held it in the past few years but since had moved to larger Airbus aircraft. Our sample was recruited through pilot forums, and open to pilots of any level of experience, First Officer or Captain. This is appropriate since pilots are trained and kept current with a homogeneous skill set across a given type of aircraft. Thus, all pilots are similarly skills-equipped to handle the scenarios we presented to them.

\subsection{Protocol}
For the purposes of control, we used the same weather conditions, traffic, and route for four runs. Pilots were asked to fly between two international airports, cruising at \num{12000}~ft, for a total flight time of around 30 minutes. Since the setup was single-pilot, the experimenter took the role of a limited co-pilot, and in doing so partly an air traffic controller. In this, the experimenter was capable of providing support in flying the aircraft but did not give input on decisions.

Each pilot was given the first run as a familiarisation flight, in which they could get used to the controls of the simulator. The following three runs included some form of attack, and was followed by a short debrief interview (provided in supplementary material). The interview assessed the pilot response to the attack, focussing on perception of impact, trust, workload and safety. We recorded data from the simulator to correlate with interview responses.

Participants knew that they were taking part in a study looking at cyber attacks on avionic systems, but did not know about the timing or type of attack. The details of the attacks were explained by the experimenter in the debrief interview.

Since the interview was conducted by the experimenter, we acknowledge that this may bias results to be more positive than if we had conducted this anonymously. This is mostly relevant to interview questions on the effectiveness of this approach as training, and we note this where appropriate.

\subsection{Equipment}

Our hardware consisted of two high-end gaming PCs, running X-Plane 11 and an aftermarket Airbus A330 model as no reliable A320 models were available, seen in Figure~\ref{fig:setup-pic}~ \cite{Xplane18}. We checked the model fidelity with type-rated Airbus A320 pilots to ensure sufficient similarity to an A320. We provided non-type-specific hardware controls, since the majority of flying on such an airliner involves manipulating automatic flight, rather than directly flying with manual controls. Participant opinions on the equipment are presented in Sec.~\ref{sec:perceptions}.

\section{Results}
\label{sec:results}

We now discuss the data collected from simulator scenarios and participant interviews. Interview response data for all scenarios can be seen in Tab.~\ref{tab:yesno} and Fig.~\ref{fig:interview}, with full data in App.~\ref{app:interview}. Responses are on the following scales:
\begin{itemize}
\setlength\itemsep{0.1em}
\item \textit{Q1.} \textbf{Confidence} in the response being the correct one, on a scale from 1, \textit{very confident}, to 5, \textit{very unconfident}.
\item \textit{Q2.} \textbf{Workload} due to the attack, on a scale from 1, \textit{no increase}, to 3, \textit{significant increase}.
\item \textit{Q3.} \textbf{Trust} in systems affect due to the attack, on a scale from 1, \textit{much more trust}, to 5, \textit{much distrust}.
\item \textit{Q4.} \textbf{Impact} on the flight due to the attack, on a scale from 1, \textit{significant impact}, to 4, \textit{no impact}.
\end{itemize}

We also recorded \textit{yes/no} responses for the following:
\begin{itemize}
\setlength\itemsep{0.1em}
\item \textit{Q5.} Whether they would trust systems under attack later in flight, N/A if they did not respond to the attack.
\item \textit{Q6.} If participants felt the attack put the aircraft in a less safe situation.
\item \textit{Q7.} If participants would respond the same way in a real aircraft (i.e. free of simulation restrictions).
\end{itemize}

\subsection{GPWS Attack}

First, we look at the GPWS scenario. We assess participants primarily on their actions, i.e. go-around, land, switch GPWS off, before considering their scale responses. 

\begin{table}[]
\centering
\caption{Action taken during GPWS attack. If a participant lands, they are not included in the numbers of following approach. Percentages are of participants in that approach.}
\label{tab:gpws-actions}
\footnotesize
\begin{tabular}{l l ll l }
\toprule
           &     & \multicolumn{2}{c}{Action Count} &       \\ \cmidrule{3-4}
 Approach           & Action    & \# & \%    & \# Participants       \\ \midrule
\cellcolor[gray]{0.85}                          &  \cellcolor[gray]{1.0}  Land      &  \cellcolor[gray]{1.0}  10 &  \cellcolor[gray]{1.0} 33.3   &  \cellcolor[gray]{0.85}                     \\
\cellcolor[gray]{0.85} \multirow{-2}{*}{1}      &  \cellcolor[gray]{0.85} Go-around &  \cellcolor[gray]{0.85} 20 &  \cellcolor[gray]{0.85} 66.7  &  \cellcolor[gray]{0.85} \multirow{-2}{*}{30}                    \\ 
\cellcolor[gray]{1.0}                           &  \cellcolor[gray]{1.0}  Turn off  &  \cellcolor[gray]{1.0}  11 &  \cellcolor[gray]{1.0} 55.0   &  \cellcolor[gray]{1.0}                          \\
\cellcolor[gray]{1.0}                           &  \cellcolor[gray]{0.85} Land      &  \cellcolor[gray]{0.85} 8  &  \cellcolor[gray]{0.85} 40.0  &  \cellcolor[gray]{1.0}                     \\
\cellcolor[gray]{1.0} \multirow{-3}{*}{2}       &  \cellcolor[gray]{1.0}  Go-around &  \cellcolor[gray]{1.0}  1  &  \cellcolor[gray]{1.0} 5.0    &  \cellcolor[gray]{1.0} \multirow{-3}{*}{20}                    \\ 
\greyrow ~3                         & ~Turn off  & ~1  & ~100.0 & ~1           

\\ \bottomrule
\end{tabular}
\end{table}

\paragraph*{Response}

\begin{figure}[]
\centering
\includegraphics[width=0.95\columnwidth, trim=15mm 218mm 110mm 18mm,clip]{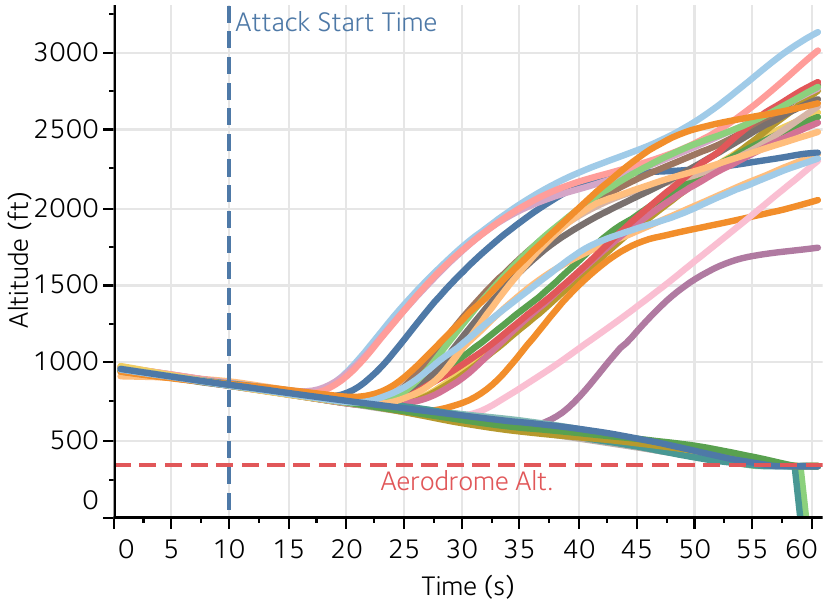}
\caption{Plot of time against altitude for first approach under GPWS attack. Each line is a participant. Eight land and disregard the alarm, on account of being sure of their position.}
\label{fig:gpws-first}

\end{figure}

Participants generally responded as expected, with a split between those opting for a terrain avoidance manoeuvre, thus a missed approach, and those disregarding the warning in order to land. The first approach is plotted in Fig.~\ref{fig:gpws-first}. In Tab.~\ref{tab:gpws-actions}, we can see that two thirds of participants went around on the first approach as a result of the alarm; these participants generally remarked that their choice was an automatic one. This is crucial as it shows that an attacker who can trigger such an attack can cause arbitrary go-arounds with reasonable chance of success. In one instance, the attack triggered late; however, in debrief, the participant noted that they would have had the same course of action regardless.

\begin{figure}
\centering
\includegraphics[width=\columnwidth,trim=15mm 259mm 110mm 18mm,clip]{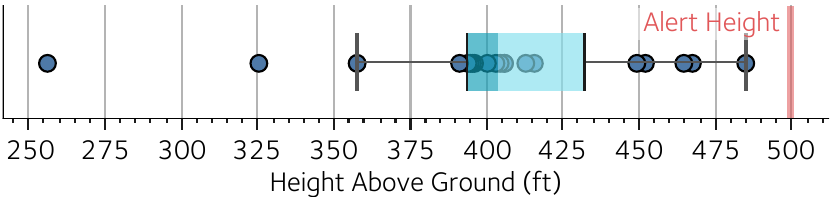}
\caption{Minimum heights reached by participants opting to go around in first approach of the GPWS attack.}
\label{fig:gpws-box}
\end{figure}

On the first approach, we found that for those opting to go around, the mean height at which the go-around began was 403.9\,ft, with a standard deviation of 51.1\,ft (see Fig.~\ref{fig:gpws-box}). Some outliers in the form of later responses do exist, shown in Fig.~\ref{fig:gpws-first}. Most participants responded within 100 ft of the alarm with an interquartile range of 29.7 ft. This is expected as pilots follow the terrain warning and execute a well-drilled manoeuvre, not allowing the aircraft to become unsafe.

To handle the attack, 11 participants switched the system off on a second approach, finding it distracting, with another one doing so on the first approach and going on to land. An attack causing GPWS to be switched off has the potential for further erosion of safety---indeed, of the 12 who switched it off, none would trust the system later in the flight. Most participants sought to silence the alarm on the second approach, though at this point they were sure of their position. 

\paragraph*{Perception}
As seen in Tab.~\ref{tab:yesno}, 14 (46.7\%) participants felt that this attack put the aircraft in a less safe situation. The numbers are lower compared to other attacks as the response is in itself a safety manoeuvre, though some pilots felt that due to its extreme nature, the aircraft is at additional risk. 

Fig.~\ref{fig:interview} shows that this scenario has the least impact as assessed by the participants---even so, it was judged to have `some impact' on average, with 8 (26.7\%) saying it was `significant'. For workload, there was on average `some increase' with 13 (43.3\%) feeling there was a `significant' increase. On top of this, a number of remarks were made about the startle factor involved on what appeared to be a normal approach. Trust in the system was eroded during the scenario, which matches with our assessment in Tab.~\ref{tab:yesno}; 29 (96.7\%) participants felt at least `some distrust'.

Generally, confidence in response was very high, with an average score of `very confident'. The majority of participants (27, 90\%) said they would take the same course of action in a real aircraft. Those who did not feel this way suggested they might have opted for a missed approach rather than landing.

\paragraph*{Evaluation} 
The reaction was fairly well defined, seemingly due to strict procedure on how to handle terrain alarms. With two thirds of participants abandoning the approach on the first alarm, an attacker can readily cause disruption. However, similar tactics on following approaches were ineffective. 
 
Clearly, a terrain avoidance manoeuvre  is the correct course of action for safety when in doubt, but if caused by an attack (and absence of dangerous terrain) rather than genuine danger, it unduly adds workload and seeds distrust in the systems. This is reinforced by all but one participant feeling some distrust, and almost half seeing a significant workload increase. 

In turn, this means that whilst some safety erosion might occur as a secondary effect, the primary impact is a repeatable disruption. A major challenge with the attack is that it cannot be easily defended against due to its short burst nature and exploitation of basic radar. Even if identified as a problem and incoming aircraft are informed about it, they would have to ignore a key safety system due to a security issue.

\subsection{TCAS Attack}

Next, we consider the TCAS attack. Results indicate that this is the most concerning attack to the participants.

\paragraph*{Response} An action summary is given in Tab.~\ref{tab:tcas-response}. We provide the `end-state' for the selected TCAS mode (e.g., if a participant selects \textit{TA Only} then \textit{Standby}, they are under \textit{Standby}) against actions taken outside of normal flight. Actions are categorised into \textit{continue on route}, i.e. no extra action taken, \textit{avoidance manoeuvre}, in which the participant changes course, or \textit{divert to origin}, i.e. return to the departure airport. Some 26 participants (87\%) turned TCAS to TA Only, with 11 (37\%) switching it to Standby. Participants switched to TA Only with a mean 4.5 RAs (standard deviation 1.7), then down to standby after another mean 2.8 TAs (standard deviation 2.1). Two participants went straight from TA/RA to Standby, one after three RAs, another after six.

Reasons cited for these changes were additional workload of responding to TAs and RAs, but also the distraction factor of repeatedly having to respond to the alerts. Looking at the control response in more detail, three of those eventually turning the transponder to TA Only and three of those turning it to standby took avoiding action. The action itself varied per participant but for some involved climbing above the planned cruise altitude or making horizontal manoeuvres to try to avoid the attacker's traffic. Two other participants diverted back to the origin airport rather than continue with malfunctioning TCAS. Three of the remaining participants did feel that TCAS was providing spurious returns but felt the risk of downgrading the system was too high and instead opted to follow the RAs as issued, rather than turn the transponder to TA Only. The final participant was not aware of the ability to go to TA Only and so remained in TA/RA. 

\begin{table}[]
\centering
\footnotesize
\caption[Participant response to the TCAS attack scenario.]{Responses to the TCAS attack scenario, mapping the final selected TCAS mode against actions or manoeuvres taken by the pilot. Percentages are of all participants.}

\label{tab:tcas-response}
\begin{tabular}{l ll ll ll ll}
\toprule
    & \multicolumn{6}{c}{Final Selected TCAS Mode}                                   &  \\ \cmidrule{2-7}
                    & \multicolumn{2}{c}{TA/RA} & \multicolumn{2}{c}{TA-Only} & \multicolumn{2}{c}{Standby} & \multicolumn{2}{c}{Total}     \\ 
\cmidrule(rl){2-3}\cmidrule(rl){4-5} \cmidrule(rl){6-7}\cmidrule(rl){8-9}
Action & \# & \% & \# & \% & \# & \% & \# & \% \\ \midrule
\greyrow \begin{tabular}[c]{@{}l@{}}Continue\\on route\end{tabular}

& 4   & 13.3 
& 10  & 33.3 
& 8   & 26.7 
& 22  & 73.3 \\  

 \begin{tabular}[c]{@{}l@{}}Avoidance\\manoeuvre\end{tabular}         
&  0 & 0.0  
&  3 & 10.0  
&  3 & 10.0  
&  6 & 20.0 \\ 

\greyrow \begin{tabular}[c]{@{}l@{}}Divert\\to origin\end{tabular}                                                              
&  0 & 0.0
&  2 & 6.7
&  0 & 0.0
&  2 & 6.7\\ \midrule
                                                                          
 Total 
&  4  & 13.3  
&  15 & 50.0  
&  11 & 36.7  
&  30 & 100.0 
\\ \bottomrule                   
\end{tabular}

\end{table}

\paragraph*{Perception} Assessing the impact, 27 (90\%) pilots felt that the attack had at least `some impact', with 19 (63\%) feeling that it had `significant impact'. This was coupled with 29 (97\%) feeling that there was at least `some increase' in workload. Typically, this increase in workload was due to having to respond to regular RAs and dealing with periodic distraction. An unduly increased workload creates further problems for the crew managing the situation, and can lead to errors.

Looking to perceived safety, 28 (97\%) of pilots felt that the attack put the aircraft in an unsafe---or potentially unsafe---situation. One cause for concern was for those on board, who might be moving about the cabin, thus injured in an extreme manoeuvre such as an RA. Some noted the possible effects on surrounding traffic in the event of following a spurious RA. Similarly, 29 (97\%) of participants felt that they had at least `some distrust' in TCAS during the scenario. Again, this is problematic as it indicates that an attacker with moderate ability can sow distrust in critical aircraft safety systems.

\begin{table}
\centering
\caption[Summary of participant actions and responses to yes/no questions as part of the debrief interview.]{Summary of participant actions and responses to debrief yes/no questions. For some participants, the question was not applicable due to previous actions, hence N/A. Percentages are of all participants, for each question.}
\label{tab:yesno}
\footnotesize
\begin{tabular}{l l ll ll ll}
\toprule
          &   & \multicolumn{6}{c}{Response}             \\ \cmidrule{3-8}
          &             & \multicolumn{2}{c}{Yes}          & \multicolumn{2}{c}{No}          & \multicolumn{2}{c}{N/A}         \\ \cmidrule(rl){3-4}\cmidrule(rl){5-6}\cmidrule(rl){7-8}

Attack & Question & \# & \%  & \# & \% & \# & \% 
\\ \midrule
\cellcolor[gray]{0.85}                              & \cellcolor[gray]{0.85} Q5--Trust                                    & \cellcolor[gray]{0.85} 1  & \cellcolor[gray]{0.85} 3.3     & \cellcolor[gray]{0.85} 25 & \cellcolor[gray]{0.85} 83.4  & \cellcolor[gray]{0.85} 4  & \cellcolor[gray]{0.85} 13.3  \\
\cellcolor[gray]{0.85}                              & \cellcolor[gray]{1.0} Q6--Safety                                      & \cellcolor[gray]{1.0} 19 & \cellcolor[gray]{1.0} 63.3    & \cellcolor[gray]{1.0} 11 & \cellcolor[gray]{1.0} 36.7  & \cellcolor[gray]{1.0} - &  \cellcolor[gray]{1.0} -        \\
\cellcolor[gray]{0.85} \multirow{-3}{*}{GS}         & \cellcolor[gray]{0.85} Q7--Same                                & \cellcolor[gray]{0.85} 28 & \cellcolor[gray]{0.85} 93.3    & \cellcolor[gray]{0.85} 2  & \cellcolor[gray]{0.85} 6.7   & \cellcolor[gray]{0.85} - &  \cellcolor[gray]{0.85} -        \\ 
                              & \cellcolor[gray]{1.0}  Q5--Trust                                    & \cellcolor[gray]{1.0}  4  & \cellcolor[gray]{1.0}  13.3    & \cellcolor[gray]{1.0}  22 &\cellcolor[gray]{1.0}   73.4  & \cellcolor[gray]{1.0}  4  &\cellcolor[gray]{1.0}   13.3  \\
                              & \cellcolor[gray]{0.85} Q6--Safety                                      & \cellcolor[gray]{0.85} 28 & \cellcolor[gray]{0.85} 93.3    & \cellcolor[gray]{0.85} 2  &\cellcolor[gray]{0.85}  6.7  & \cellcolor[gray]{0.85} - & \cellcolor[gray]{0.85}  -        \\
\multirow{-3}{*}{TCAS}        & \cellcolor[gray]{1.0}  Q7--Same                                & \cellcolor[gray]{1.0}  30 & \cellcolor[gray]{1.0}  100.0   & \cellcolor[gray]{1.0}  0  &\cellcolor[gray]{1.0}   0.0   & \cellcolor[gray]{1.0}  - & \cellcolor[gray]{1.0}   -        \\ 
\cellcolor[gray]{0.85}                              & \cellcolor[gray]{0.85} Q5--Trust                                    & \cellcolor[gray]{0.85} 0  & \cellcolor[gray]{0.85} 0.0     & \cellcolor[gray]{0.85} 12 &\cellcolor[gray]{0.85}  40.0  & \cellcolor[gray]{0.85} 18 &\cellcolor[gray]{0.85}  60.0 \\
\cellcolor[gray]{0.85}                              & \cellcolor[gray]{1.0}  Q6--Safety                                      & \cellcolor[gray]{1.0}  14 & \cellcolor[gray]{1.0}  46.7    & \cellcolor[gray]{1.0}  16 &\cellcolor[gray]{1.0}   53.3  & \cellcolor[gray]{1.0}  - & \cellcolor[gray]{1.0}   -        \\
\cellcolor[gray]{0.85}\multirow{-3}{*}{GPWS}        & \cellcolor[gray]{0.85} Q7--Same                                & \cellcolor[gray]{0.85} 27 & \cellcolor[gray]{0.85} 90.0    & \cellcolor[gray]{0.85} 3 & \cellcolor[gray]{0.85} 10.0   & \cellcolor[gray]{0.85} - & \cellcolor[gray]{0.85}  -       \\ \bottomrule
\end{tabular}

\end{table}

\paragraph*{Evaluation} In this scenario, the most common option was to reduce the alerting level of TCAS to either only notify of traffic (TA Only) or to switch the system off. We were particularly interested in how many RAs would cause a pilot to reduce this sensitivity, since this is the point at which an attacker can reduce the effect of a safety system. On average, our participants took 4-5 RAs to reduce the sensitivity, or approximately 7 to switch it off. Given that TCAS is a final line of defence against mid-air collision, this is problematic.

This attack presents the best example of a safety against security trade-off. If TCAS is left at full sensitivity, having the attacker able to trigger extreme changes of course through RAs puts other aircraft at risk of also having TCAS alarms. Furthermore, resolving these RAs adds workload for the crew and potentially removes their attention from other tasks. 

Accounting for this, switching the system to TA Only is a fair balance as it does not lose all the situational awareness that TCAS provides. However, disruption is caused before this point---repeated RAs would affect nearby traffic, passenger comfort and safety as well as crew workload.

\subsection{Glideslope Spoof}

We now look at the glideslope spoof, where an attacker aims to capture a pilot on a false GS. We focus on the first approach, in which the participants knew least about the attack.

\paragraph*{Response}

On encountering the attack, 4 (13.3\%) participants chose to land anyway on account of having a good visual picture. Of the 26 (86.7\%) participants choosing to go around, three went around a further time. The choices after participants identified a problem with the GS are as follows: 
\begin{itemize}
\setlength\itemsep{0.1em}
\item 1 (3.3\%) used a VHF Omnidirectional Range approach, 
\item 2 (6.7\%) used a Surveillance Radar Approach (SRA), which relies on higher involvement with ATC,
\item 8 (26.7\%) flew a localizer only approach (LOC DME) on account of identifying GS problems,
\item 9 (30.0\%) dropped ILS completely, and used an Area Navigation (RNAV) approach, which is based on GPS,
\item 6 (20.0\%) flew a visual approach due to good conditions.
\end{itemize}

The split highlights that the attack invokes a response grey area. Eleven participants  chose to forgo ILS completely and use SRA or RNAV approaches as they could not identify the issue. However, eight were happy to use LOC DME since they had identified that just the GS was affected. 

\begin{figure}[t]
\subfloat[Height above ground level at point of first go-around.]{%
	\includegraphics[width=0.98\columnwidth,trim=15mm 257mm 110mm 18mm,clip]{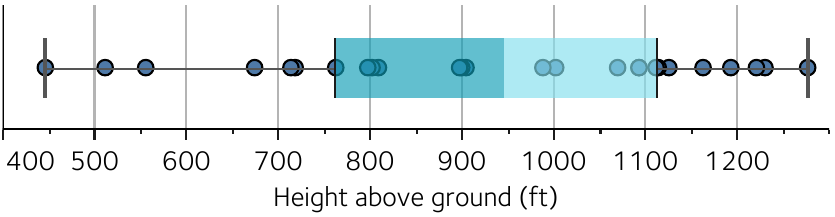}
	\label{fig:glideslope-height-box}
}
\\
\subfloat[Distance from runway touchdown zone at first go-around.]{%
	\includegraphics[width=0.98\columnwidth,trim=15mm 258mm 110mm 18mm,clip]{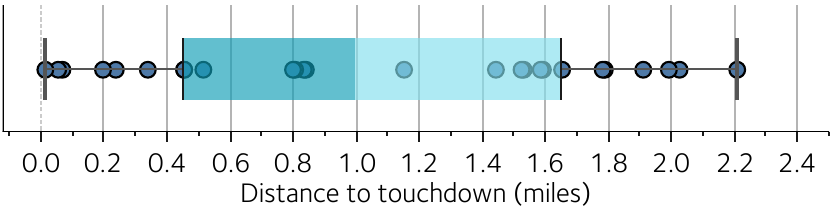}
	\label{fig:glideslope-dist-box}
}
\caption{Box plots of participants performing a go-around on the first approach under the glideslope attack.}

\end{figure}

Fig.~\ref{fig:glideslope-height-box} and \ref{fig:glideslope-dist-box} show box plots for the height above ground level (in feet) and distance (in miles) from the runway touchdown zone, respectively. This is plotted for each participant opting to go-around on the first approach at the point of choosing to abort the approach. The mean go-around altitude was 930.0 ft, with a standard deviation of 235.8 ft, and distance of 1.1 miles, with a standard deviation of 0.7 miles. Since preparation for a go-around takes a few seconds, the mean point is just as participants descended below 1000\,ft. 

Considering that a \ang{3} GS has a rate of descent of 700\,ft/min, this means the go-around begins with just over a minute to touchdown. In poor weather, this might be the first time the pilots see the runway, making for a short amount of time to abort the approach. That the attack is subtle enough for the aircraft to get so close to landing demonstrates how difficult it is to clearly identify that an ILS attack is under way.

\paragraph*{Perception}
Some 13 (43.3\%) participants found the attack had `some' impact or greater, as shown in Fig.~\ref{fig:interview}. The GS attack had a small perceived workload increase with 22 (73.3\%) participants claiming `some' increase, which may be due to the GS attack occurring gradually, higher above the ground with PAPIs providing visual checks. A number of participants noted that this attack would be harder to deal with in worse weather conditions. Even so, 26 (86.7\%) participants performed a go-around as a result of being unsure.

As with TCAS and GPWS, the attack caused `some' distrust in aircraft systems, with 23 (76.7\%) participants remarking `some' or `significant' impact. However, some participants correctly identified that the ground systems were at fault and so did not distrust the aircraft. Furthermore, Tab.~\ref{tab:yesno} shows that of the 26 (86.7\%) participants, who did perform a go-around, all but one would not trust the GS on a second approach. This is reinforced by the fact that 19 (63.3\%) felt that the attack put the aircraft in a less safe situation.

\paragraph*{Evaluation} 
As well as almost two thirds of participants feeling that the attack put the aircraft in a less safe situation, over 85\% of participants abort their first approach and go around. Although subtle, this attack has potential for disruption. Even though some participants did not identify the problem, the indication of one was enough for the majority to avoid the GS and choose another approach method. Switching to a different approach not only increases workload---as indicated in our perception responses---but may affect nearby traffic. For example, an SRA approach requires heavy ATC interaction, tying up a controller temporarily. Furthermore, not all airports are equipped to handle all approach types. 

Such a division is interesting; whilst the most safe and secure option is to use an approach outside of the ILS such as SRA or RNAV, no single approach emerged as a favourite. This could be due to the GS not malfunctioning in expected ways such as bending or being too steep. On top of this, participants either responded quite late, i.e. whilst over the runway, or quite early in choosing to go around (Fig.~\ref{fig:glideslope-dist-box}. The group of late responders indicates the subtlety of the attack as they waited to see what happened; in this case, an early abort and switch to a different approach is the safer choice.

Attempting to perform this attack on numerous consecutive aircraft would likely see ATC instructing to not use the ILS, making the attack a form of denial of service. Such an eventuality has a range of possible consequences including diversions or reduced airport capacity due to higher separation.

\section{Discussion}
\label{sec:discuss}

We now explore the effects across all considered attacks, how they affect the way the aircraft is flown and on the participants.

\subsection{Effect on Safety}

On balance, our expected response for each attack was for participants to take the `safest' option in the circumstances which typically led to disruption. Because of this, we did not necessarily expect much perceived loss of safety. This was not the case; for TCAS and GS, 93.3\% and 63.3\% felt the attack made the aircraft less safe. For TCAS, this is likely due to the uncertainty of the situation, with pilots not expecting false alarms. In the case of GS, the safety concern comes from how late the discrepancy is apparent and the situation this leaves the aircraft in. The exception to this was the GPWS attack, in which the terrain avoidance manoeuvre is the de facto safe option so fewer pilots felt safety was affected. Arguably, this highlights the interplay of safety and security the most. Even though most pilots took the safe option, they still felt they were compromised by factors out of their control. 

Furthermore, the range of decisions taken indicates that these attacks do indeed create gray areas for responses. In an industry where safety is paramount and relies on well-defined procedure, the ability for an attacker to create situations open for interpretation is concerning. It is especially so when considering that these are systems tied closely to aircraft safety.

\subsection{Cost of Disruption}
We have demonstrated the ability for these attacks to cause missed approaches and diversions. With this in mind, we can estimate the cost of this attack on an aircraft. As an example, we use Boeing aircraft due to public fuel usage information to calculate the cost of a missed approach. For their smaller 737-800 aircraft, the missed approach uses 127\,kg (41.79\,gal) more fuel than a successful one; for the larger 777-200, it is 399\,kg (111.55\,gal) more~\cite{Roberson2010}. Coupled with a nominal jet fuel cost of 184.58\,c/gal, this costs approximately \$77 for the 737 or \$205 for the 777.\footnote{Calculated using IATA Jet Fuel Price Monitor for $18^{th}$ January 2019~\cite{IATA2018}.} Added to the expense of further time in the air---more difficult to predict as it depends on factors such as the airfield and traffic---plus a second approach, which costs approximately \$139 (using 230\,kg, or 75.68\,gal) or for the 737, or \$516 for the 777 (using 850\,kg, or 279.69\,gal), this becomes expensive for the airline. 

Diversions add further expense, with potential effects on scheduling or passenger inconvenience. The UK Civil Aviation Authority estimates that these can cost an airline between \pounds\num{10000}--\pounds\num{80000}, depending on the size of the aircraft and location of diversion~\cite{CAA2018}. For example, passenger disruption causing diversion aboard a Norweigan flight cost \euro\num{100000} in 2018~\cite{Dorsey2017,IrishExaminer2015}. Closed airports are similarly costly, with drones closing London Gatwick for two days in December 2018 and costing airline Easyjet \pounds 15\,million~\cite{Kollewe2019}.

\subsection{Additional Impact Factors}

In debrief, participants raised a number of other factors which would affect the impact of attacks. Weather conditions were prominent; all scenarios would be more difficult to handle in poor visibility. Particularly for the approach-based GS and GPWS attacks, good visibility allowed participants to arrive at their decisions more quickly. Some participants noted it would be hard to identify the GS attack under automatic landing conditions, leaving much less time for pilots to respond. 

Other contributing factors include tiredness and terrain. In response to the GPWS attack, one participant who chose not to go around commented that their action in a real aircraft would depend on tiredness, as well as weather and how busy the crew were. Again in the GPWS attack, others identified that terrain surrounding the airport affects their choice---they would be much more likely to abort an approach in challenging terrain, and less if they are familiar with the airport.

\subsection{Simulation for Training}
\label{sec:perceptions}
To assess whether responses were realistic, we asked each participant whether their response to each scenario would be the same in a real aircraft. We found that for:
\begin{itemize}
\setlength\itemsep{0.1em}
\item \textbf{GPWS}, 27 (90.0\%) would do the same, and the remaining three would go around in the same scenario again,
\item \textbf{TCAS}, 30 (100.0\%) would do the same,
\item \textbf{Glideslope}, 28 (93.3\%) would do the same with the remaining two opting to go around and revert to RNAV.
\end{itemize}

We asked each participant for their views the value of such experiments or training in preparation for cyber attack. All participants felt the scenarios were useful, and 28 (93.3\%) commented that training for cyber attacks using a simulator would be valuable. We also asked if they felt limited by the simulation set up, on a scale of `1--\textit{not}', `2--\textit{somewhat}' and `3--\textit{heavily}' limited, with the average response corresponding to `somewhat'. The main limits were lack of a second crew member, and the general (rather than Airbus specific) controls. We do note that these figures are subject to some bias but we feel that the results are sufficiently strong that the effect on our conclusion is minimal. 

The results suggest that this method can be valuable both in identifying crew response to attacks and providing cyber attack readiness. Furthermore, the fact that the scenarios in this paper lie in procedural gray areas and do not have a series of steps to resolve them provides an ideal opportunity for training. One point of caution is negative training, with some participants noting that care must be taken to avoid training pilots to ignore or distrust their systems.

\section{Mitigations}
\label{sec:mitigate}
Most of the security problems enabling these attacks are due to a lack of deployed security mechanisms. This section offers potential mitigations, outside of expensive redeployment.

\paragraph*{Spectrum Monitoring} Both GPWS and GS attacks require the attacker to transmit at a reasonable power level near to the airport. Early warning of such attacks could be provided using spectrum monitoring around the airfield on key aviation frequencies, compared against fingerprints of legitimate signals and regions in which they should be transmitted. This could be achieved with low-cost sensor networks, however a reasonable investment into infrastructure to process and store the collected data would be needed.

\paragraph*{Crowdsourced Air Traffic Surveillance} A larger scale version of this would use  crowdsourced networks collecting air traffic surveillance signals to detect message injection. This would be useful against the TCAS since it can occur over a much larger area. These networks, such as Flightradar24~\cite{Flightradar24AB2018} or Opensky Network~\cite{Schafer2014}, consist of geographically distributed low-cost sensors operated by members of the public, who feed data to a central server. A benefit of this kind of network is the ability to cross-check legitimate message reception, helping to identify and locate rogue messages. A theoretical foundation for using this can be seen in~\cite{Strohmeier2017cycon}.

\paragraph*{Training for Pilot Awareness} As demonstrated, the participants handled the scenarios safely; existing training provides the skills to do so. However, pilots also felt that the situations led to losses in safety, and that the scenarios had value from exposing them to unusual situations. Naturally, processes which help reduce startling and give experience in handling new situations can be beneficial (see Sec.~\ref{sec:relwork}). To develop this into a training exercise, more complete and realistic scenarios incorporating a full crew and realistic ATC could be developed.

\paragraph*{Security by Design} Longer term, avionic communications systems need secure design. Current trends suggest that next generation data links, likely to come into use in around 10 years, have some security by default. AeroMACS is the most developed example so far and derivative of IEEE 802.16e (WiMAX), from which it inherits secure communications~\cite{AeroMACS2018,IEEE2005,SESAR2014}. However, the systems in this paper will not be replaced soon and security needs to be patched in. Both TCAS and GS/ILS have the opportunity to make such changes. For ILS signals, research looking into using Distance Measuring Equipment (DME) for data carrying could provide a second channel for ILS integrity and authentication~\cite{Zeng2018}. Such work for TCAS does not yet exist.

\section{Conclusion}
\label{sec:conclude}
In this paper, we describe three novel practical wireless interference attacks on avionics systems. We implement them in a standard flight simulator and test their impact with 30 pilots in the loop. The attacks are not sufficiently mitigated by aviation's prevalent safety culture and can cause disruption, financial loss, and reputational damage. Crucially, participants saw a reduction in safety under the wrong circumstances, e.g. as safety systems were switched off in 38\% of cases. 

Our results imply that the attack on TCAS is the most concerning as it combines widespread inconvenience and potential safety reduction. Both GS/ILS and GPWS also pose problems, though are easier to mitigate on the flight deck. Finally, we conclude that flight simulation for wireless attack awareness or training has potential to aid and prepare crew. Since preventative security by design will not be deployable in the near-term, such training could be highly valuable.

\bibliographystyle{plain}
\bibliography{refs}

\begin{thebibliography}{10}

\bibitem{AAIPC2006}
{Final Report A-00X{\/}CENIPA{\/}2008}.
\newblock Technical report, {Aeronautical Accident Investigation and Prevention
  Center}, September 2006.

\bibitem{AirLinePilotsAssociationIntl2017}
{Airline Pilots Association}.
\newblock {Aviation Cyber Security: The Pilot's perspective}.
\newblock Technical report, Air Line Pilots Association Int'l, Washington,
  2017.

\bibitem{ARINC2015}
{ARINC}.
\newblock {ARINC Characteristic 735B-2: Traffic Computer TCAS and ADS-B
  Functionality}.
\newblock Technical Report 735B-2, 2015.

\bibitem{Beeb2018}
BBC.
\newblock {2017 safest year for air travel as fatalities fall}.
\newblock \url{https://www.bbc.com/news/business-42538053}, January 2018.
\newblock Accessed on 2018-11-20.

\bibitem{Breen2015}
Barry~C. Breen.
\newblock {\em {Digital Avionics Handbook}}, chapter~21, pages 21.1--21.12.
\newblock CRC Press, 3rd edition, 2015.
\newblock gpws chapter.

\bibitem{Buch2019}
Jan-Philipp Buch, Robert~Manuel Geister, Luca Canzian, Giovanni Gamba, and
  Oscar Pozzobon.
\newblock {What the Hack Happened to the Flight Deck: Analyzing the Impact of
  Cyberattacks on Commercial Flight Crews}.
\newblock In {\em AIAA SciTech 2019}, January 2019.

\bibitem{Casner2013}
Stephen~M Casner, Richard~W Geven, and Kent~T Williams.
\newblock The effectiveness of airline pilot training for abnormal events.
\newblock {\em Human factors}, 55(3):477--485, 2013.

\bibitem{CAA2018}
{Civil Aviation Authority}.
\newblock {Disruptive Passengers}.
\newblock
  \url{https://www.caa.co.uk/Passengers/On-board/Disruptive-passengers/}, 2018.
\newblock Accessed on 2018-11-20.

\bibitem{Costin2012}
Andrei Costin and Aur{\'{e}}lien Francillon.
\newblock {Ghost in the Air(Traffic): On insecurity of ADS-B protocol and
  practical attacks on ADS-B devices}.
\newblock In {\em Black Hat USA}, pages 1--10, jul 2012.

\bibitem{Cox2018}
Joseph Cox.
\newblock {US Government Probes Airplane Vulnerabilities, Says Airline Hack Is
  `Only a Matter of Time'}.
\newblock
  \url{https://motherboard.vice.com/en_us/article/d3kwzx/documents-us-government-hacking-planes-dhs},
  June 2018.
\newblock Accessed on 2019-02-15.

\bibitem{Dahlstrom2009}
N~Dahlstrom, Sidney Dekker, R~Van~Winsen, and J~Nyce.
\newblock Fidelity and validity of simulator training.
\newblock {\em Theoretical Issues in Ergonomics Science}, 10(4):305--314, 2009.

\bibitem{Dorsey2017}
Brendan Dorsey.
\newblock {Hawaiian Airlines Passenger Fined \$100,000 for Bad Behavior}.
\newblock
  \url{https://thepointsguy.com/2017/08/hawaiian-airlines-passenger-fined/},
  August 2017.
\newblock Accessed on 2018-11-20.

\bibitem{Eurocontrol2014transponder}
Eurocontrol.
\newblock Flying without a transponder---10 minutes is all it can take.
\newblock {\em NetAlert}, (19):5, May 2014.

\bibitem{Eurocontrol2014failure}
Eurocontrol.
\newblock Transponder failure is not always total.
\newblock {\em NetAlert}, (19):6--7, May 2014.

\bibitem{EASA2018}
{European Aviation Safety Agency}.
\newblock {{Impact Assessment of Cybersecurity Threats}}.
\newblock Technical Report {EASA\_REP\_RESEA\_2016\_1}, {European Union}, 2018.

\bibitem{IrishExaminer2015}
Irish Examiner.
\newblock {Limerick court fines man \euro 1,000 after disrupting flight to tune
  of \euro 100k}.
\newblock
  \url{https://www.irishexaminer.com/breakingnews/ireland/limerick-court-fines-man-1000-after-disrupting-flight-to-tune-of-100k-674943.html},
  April 2015.
\newblock Accessed on 2018-11-20.

\bibitem{FAA2011a}
{Federal Aviation Administration}.
\newblock {\em {Introduction to TCAS II Version 7.1}}, chapter~1, pages
  {5--10}.
\newblock {U.S. Department of Transport}, 2011.

\bibitem{FAA2011TCASModeS}
{Federal Aviation Administration}.
\newblock {\em {Introduction to TCAS II Version 7.1}}, chapter~1, page~{17}.
\newblock {U.S. Department of Transport}, 2011.

\bibitem{FAA2011TCASModeC}
{Federal Aviation Administration}.
\newblock {\em {Introduction to TCAS II Version 7.1}}, chapter~1, pages
  {17--19}.
\newblock {U.S. Department of Transport}, 2011.

\bibitem{FAA2011TCASTau}
{Federal Aviation Administration}.
\newblock {\em {Introduction to TCAS II Version 7.1}}, chapter~1, pages
  {22--24}.
\newblock {U.S. Department of Transport}, 2011.

\bibitem{FAA2012}
{Federal Aviation Administration}.
\newblock {\em {Instrument Flying Handbook}}, chapter~9, pages {9.35--9.38}.
\newblock Number FAA-H-8083-15B. {U.S. Department of Transport}, 2012.

\bibitem{FAA2012ILSError}
{Federal Aviation Administration}.
\newblock {\em {Instrument Flying Handbook}}, chapter~9, page {9.40}.
\newblock Number FAA-H-8083-15B. {U.S. Department of Transport}, 2012.

\bibitem{FAA2015}
{Federal Aviation Administration}.
\newblock {Lighting Systems – Precision Approach Path Indicators (PAPI)}.
\newblock
  \url{https://www.faa.gov/about/office_org/headquarters_offices/ato/service_units/techops/navservices/lsg/papi/},
  June 2015.
\newblock Accessed on 2018-11-23.

\bibitem{Flightradar24AB2018}
{Flightradar24 AB}.
\newblock {Flightradar24 FAQs}.
\newblock \url{https://www.flightradar24.com/faq}, 2018.
\newblock Accessed on 2018-03-15.

\bibitem{Gontar2018}
Patrick Gontar, Hendrik Homans, Michelle Rostalski, Julia Behrend,
  Fr{\'e}d{\'e}ric Dehais, and Klaus Bengler.
\newblock {Are pilots prepared for a cyber-attack? A human factors approach to
  the experimental evaluation of pilots' behavior}.
\newblock {\em {Journal of Air Transport Management}}, vol. 69:pp. 26--37, June
  2018.

\bibitem{Hays1992}
Robert~T. Hays, John~W. Jacobs, Carolyn Prince, and Eduardo Salas.
\newblock Flight simulator training effectiveness: A meta-analysis.
\newblock {\em Military Psychology}, 4(2):63--74, 1992.

\bibitem{Henely2015}
Steve Henely.
\newblock {\em {Digital Avionics Handbook}}, chapter~22, pages 22.1--21.
\newblock CRC Press, 3rd edition, 2015.
\newblock {TCAS Chapter}.

\bibitem{Hovav2011}
Ron Hovav.
\newblock {Airbus Erroneus Radio Altitudes}.
\newblock Technical Report {FMG/15 – WP/08}, International Civil Aviation
  Organization, 2011.

\bibitem{IEEE2005}
{IEEE Computer Society}.
\newblock {IEEE Standard for Local and Metropolitan Area Networks Part 16: Air
  Interface for Fixed and Mobile Broadband Wireless Access Systems}.
\newblock Technical Report 802.16e-2005, 2015.

\bibitem{SESAR2014}
{INDRA}.
\newblock {AEROMACS -- Security Analysis}.
\newblock Technical Report 15.02.17, SESAR Joint Undertaking, 2014.

\bibitem{IATA2018}
{International Air Transport Association (IATA)}.
\newblock {Jet Fuel Price Monitor}.
\newblock
  \url{https://www.iata.org/publications/economics/fuel-monitor/Pages/index.aspx},
  November 2018.
\newblock Accessed on 2018-11-20.

\bibitem{ICAO2006}
{International Civil Aviation Organization}.
\newblock {\em {Annex 10 to the Convention on International Civil
  Aviation---Aeronuatical Telecommunications}}, volume~1, chapter~3, pages
  {3.19--3.20}.
\newblock 2006.

\bibitem{ICAO2018}
{International Civil Aviation Organization}.
\newblock {Radio Altimeter Spectrum}.
\newblock
  \url{https://www.icao.int/NACC/Documents/Meetings/2018/RPG/RPGITUWRC2019-P08.pdf},
  February 2018.
\newblock Reference Number RPGITUWRC2019-P08, Accessed on 2018-11-21.

\bibitem{Johnson2013}
Daniel~P Johnson.
\newblock {Civil Aviation and CyberSecurity}.
\newblock
  \url{http://sites.nationalacademies.org/cs/groups/depssite/documents/webpage/deps_084768.pdf},
  2013.
\newblock Accessed on 2019-02-04.

\bibitem{Kochan2004}
Janeen~A Kochan, Eyal~G Breiter, and Florian Jentsch.
\newblock Surprise and unexpectedness in flying: Database reviews and analyses.
\newblock In {\em Proceedings of the Human Factors and Ergonomics Society
  Annual Meeting}, volume~48, pages 335--339. SAGE Publications Sage CA: Los
  Angeles, CA, 2004.

\bibitem{Kollewe2019}
Julia Kollewe and Gwyn Topham.
\newblock {EasyJet says Gatwick drone chaos cost it \pounds15m}.
\newblock
  \url{https://www.theguardian.com/business/2019/jan/22/easyjet-gatwick-drone-cost-brexit-flights},
  January 2019.
\newblock Accessed on 2019-02-15.

\bibitem{Xplane18}
{Laminar Research}.
\newblock X-plane 11.
\newblock \url{https://www.x-plane.com/}, August 2018.
\newblock Accessed on 2018-11-21.

\bibitem{Landman2017}
Annemarie Landman, Eric~L Groen, MM~Van~Paassen, Adelbert~W Bronkhorst, and Max
  Mulder.
\newblock The influence of surprise on upset recovery performance in airline
  pilots.
\newblock {\em The International Journal of Aerospace Psychology},
  27(1-2):2--14, 2017.

\bibitem{Landman2018}
Annemarie Landman, Peter van Oorschot, M.~M.~(Rene) van Paassen, Eric~L. Groen,
  Adelbert~W. Bronkhorst, and Max Mulder.
\newblock Training pilots for unexpected events: A simulator study on the
  advantage of unpredictable and variable scenarios.
\newblock {\em Human Factors}, 60(6):793--805, 2018.

\bibitem{Miller2012}
Jordan Miller.
\newblock {Handling ILS anomalies}.
\newblock
  \url{http://www.ifr-magazine.com/issues/1_24/features/Handling-ILS-anomalies_240-1.html},
  September 2012.
\newblock Accessed on 2018-12-13.

\bibitem{Ofcom2018}
{Ofcom}.
\newblock {UK Amateur Radio License -- Terms, Conditions and Limitations}.
\newblock
  \url{https://www.ofcom.org.uk/__data/assets/pdf_file/0027/62991/amateur-terms.pdf}.

\bibitem{OpenskyNetwork2018}
{OpenSky Network}.
\newblock {dump1090 Decoder with High Precision Timestamping}.
\newblock \url{https://github.com/openskynetwork/dump1090-hptoa}, November
  2018.
\newblock Accessed on 2018-11-23.

\bibitem{Patel2012}
Raju Patel.
\newblock {Managing Cybersecurity Risk}.
\newblock
  \url{https://www.omg.org/news/meetings/tc/va-17/special-events/cybersecurity-pdf/Dr-Raju-Patel-Managing-Cybersecurity-Risk-in-Weapons-Systems-3-21-17.pdf},
  2017.
\newblock Accessed on 2019-02-03.

\bibitem{Roberson2010}
William Roberson and James~A. Johns.
\newblock {Fuel Conservation Strategies: Descent and Approach}.
\newblock {\em AERO}, (38):25--28, 2010.

\bibitem{Salas1998}
Eduardo Salas, Clint~A Bowers, and Lori Rhodenizer.
\newblock It is not how much you have but how you use it: Toward a rational use
  of simulation to support aviation training.
\newblock {\em The International Journal of Aviation Psychology},
  8(3):197--208, 1998.

\bibitem{Sanfilippo2017}
Salvatore Sanfilippo.
\newblock dump1090.
\newblock \url{https://github.com/antirez/dump1090}, January 2017.
\newblock Accessed on 2018-11-23.

\bibitem{Sathaye2019}
Harshad Sathaye, Domien Schepers, Aanjhan Ranganathan, and Guevara Noubir.
\newblock {Wireless Attacks on Aircraft Instrument Landing Systems}.
\newblock In {\em 28th USENIX Security Symposium}, Aug 2019.

\bibitem{Saul-Pooley2017gs}
Dorothy Saul-Pooley.
\newblock {\em {Radio Navigation \& Instrument Flying}}, chapter~15, pages
  325--329.
\newblock Pooley's, 2017.
\newblock Glideslope.

\bibitem{Saul-Pooley2017loc}
Dorothy Saul-Pooley.
\newblock {\em {Radio Navigation \& Instrument Flying}}, chapter~15, pages
  325--321.
\newblock Pooley's, 2017.
\newblock Localiser.

\bibitem{Schafer2013}
Matthias Sch{\"a}fer, Vincent Lenders, and Ivan Martinovic.
\newblock Experimental analysis of attacks on next generation air traffic
  communication.
\newblock In {\em International Conference on Applied Cryptography and Network
  Security}, pages 253--271. Springer, 2013.

\bibitem{Schafer2014}
Matthias Sch{\"{a}}fer, Martin Strohmeier, Vincent Lenders, Ivan Martinovic,
  and Matthias Wilhelm.
\newblock {Bringing Up OpenSky: A Large-scale ADS-B Sensor Network for
  Research}.
\newblock {\em Proceedings of the 13th International Symposium on Information
  Processing in Sensor Networks (IPSN 2014)}, pages 83--94, 2014.

\bibitem{Skybrary2018tcas}
{SKYbrary}.
\newblock {Airborne Collision Avoidance System (ACAS)}.
\newblock
  \url{https://www.skybrary.aero/index.php/Airborne_Collision_Avoidance_System_(ACAS)#Complying_with_RAs},
  2017.
\newblock Accessed on 2018-08-30.

\bibitem{Skybrary2018pullup}
{SKYbrary}.
\newblock {Response to a ``PULL UP'' Warning}.
\newblock
  \url{https://www.skybrary.aero/index.php/Response_to_a_\%22PULL_UP\%22_Warning},
  2017.
\newblock Accessed on 2018-08-30.

\bibitem{Smith2017}
Matthew Smith, Daniel Moser, Martin Strohmeier, Vincent Lenders, and Ivan
  Martinovic.
\newblock {Economy Class Crypto: Exploring Weak Cipher Usage in Avionic
  Communications via ACARS}.
\newblock In {\em 21st International Conference on Financial Cryptography and
  Data Security}, Malta, 2017.

\bibitem{Strohmeier2014a}
Martin Strohmeier, Vincent Lenders, and Ivan Martinovic.
\newblock {On the Security of the Automatic Dependent Surveillance-Broadcast
  Protocol}.
\newblock {\em IEEE Communications Surveys \& Tutorials}, 17(2):1066--1087,
  2015.

\bibitem{Strohmeier2018}
Martin Strohmeier, Anna~K Niedbala, Matthias Sch{\"a}fer, Vincent Lenders, and
  Ivan Martinovic.
\newblock {Surveying Aviation Professionals on the Security of the Air Traffic
  Control System}.
\newblock In {\em International Workshop on Cyber Security for Intelligent
  Transportation Systems (CSITS)}, September 2018.

\bibitem{Strohmeier2017}
Martin Strohmeier, Matthias Sch{\"a}fer, Rui Pinheiro, Vincent Lenders, and
  Ivan Martinovic.
\newblock {On Perception and Reality in Wireless Air Traffic Communication
  Security}.
\newblock {\em IEEE Transactions on Intelligent Transportation Systems},
  18(6):1338--1357, June 2017.

\bibitem{Strohmeier2017cycon}
Martin Strohmeier, Matthew Smith, Matthias Sch{\"a}fer, Vincent Lenders, and
  Ivan Martinovic.
\newblock {Crowdsourcing Security for Wireless Air Traffic Communications}.
\newblock In {\em {Cyber Conflict (CyCon), 2017 9th International Conference
  on}}. IEEE, May 2017.

\bibitem{Trautvetter2012}
Chad Trautvetter.
\newblock {FAA Reminds Pilots of Possible ILS Interference}.
\newblock
  \url{https://www.ainonline.com/aviation-news/business-aviation/2012-04-24/faa-reminds-pilots-possible-ils-interference},
  April 2012.
\newblock Accessed on 2018-11-23.

\bibitem{UKCAA1976}
{UK Civil Aviation Authority}.
\newblock {Ground Proximity Warning Systems}.
\newblock \url{https://publicapps.caa.co.uk/docs/33/CASPEC14.PDF}, 1976.
\newblock Accessed on 2018-12-16.

\bibitem{AeroMACS2018}
{WiMAX Forum}.
\newblock {AeroMACS}.
\newblock http://wimaxforum.org/Page/AeroMACS, 2018.
\newblock Accessed on 2018-08-31.

\bibitem{Young2018}
C.~Young.
\newblock Stratux.
\newblock \url{http://stratux.me/}, 2018.
\newblock Accessed on 2018-12-13.

\bibitem{Yusupov2017}
Linar Yusupov.
\newblock {ADSB-Out}.
\newblock \url{https://github.com/lyusupov/ADSB-Out}, December 2017.
\newblock Accessed on 2018-11-23.

\bibitem{Zeng2018}
D.~Zeng, F.~Box, J.~C. Ashley, L.~Globus, D.~V. Baraban, F.~A. Niles, and
  B.~Phillips.
\newblock {DME Potential for Data Capability}.
\newblock In {\em 2018 Integrated Communications, Navigation, Surveillance
  Conference (ICNS)}, pages 4D3--1--4D3--16, April 2018.

\end{thebibliography}

\appendix
\section{Scale Response Data}
In Figure~\ref{fig:interview}, we provide a chart representation of interview responses by participants, relating to their experiences of each attack. We provide the full data for this table on following page.
\label{app:interview}
\begin{table*}[tp]
\centering
\caption{Summary of participant interview responses for attack scenarios. Scale points are normalized so that 1 represents the most `positive' point, i.e. the greatest change, and the highest value represents the most `negative' i.e. no change. For example, using Q4 relating to impact, 1 is the `significant impact' response. Dash indicates where no scale value existed, and representative scale point is taken as scale response at the rounded mean, e.g. for impact, 1.4 will be `significant impact'.}
\begin{tabular}{l l l l l l l l l l l l l l}
\toprule
                      &            & \multicolumn{9}{c}{Number of Participant Responses per Scale Point}             &      &                                                                      &                                                      \\ \cmidrule{3-11}
Attack                                                              & Question   & 1  & 1.5 & 2  & 2.5 & 3 & 3.5 & 4  & 4.5 & 5  & Mean & \begin{tabular}{@{}l@{}}Representative\\ Scale Point\end{tabular} & Std. Dev \\ \midrule
\greyrow                                                            & Impact     & 10 & 3   & 10 & 2   & 4 & 0   & 1  & -   & -  & 1.85 & Some impact                                                          & 0.787                                                \\
\cellcolor[gray]{0.85}                                              & Confidence & 21 & 1   & 8  & 0   & 0 & 0   & 0  & 0   & 0  & 1.28 & Very confident                                                       & 0.441                                                \\
\greyrow                                                            & Workload   & 6  & 1   & 22 & 1   & 0 & -   & -  & -   & -  & 1.80 & Some increase                                                        & 0.420                                                \\
\cellcolor[gray]{0.85}~\multirow{-4}{*}{GS}                          & Trust      & 5  & 0   & 18 & 0   & 7 & 0   & 0  & 0   & 0  & 2.07 & Some distrust                                                        & 0.629                                                \\ 
\greyrow \cellcolor[gray]{1.0}                                      & Impact     & 19 & 3   & 5  & 2   & 1 & 0   & 0  & -   & -  & 1.38 & Significant impact                                                   & 0.573                                                \\
                                                                    & Confidence & 12 & 4   & 11 & 0   & 3 & 0   & 0  & 0   & 0  & 1.63 & Somewhat confident                                                   & 0.639                                                \\
\greyrow  \cellcolor[gray]{1.0}                                     & Workload   & 16 & 4   & 9  & 1   & 0 & -   & -  & -   & -  & 1.42 & Significant increase                                                 & 0.484                                                \\
~\multirow{-4}{*}{TCAS}                                             & Trust      & 19 & 2   & 8  & 0   & 1 & 0   & 0  & 0   & 0  & 1.36 & Much distrust                                                        & 0.531                                                \\ 
 \greyrow                                                           & Impact     & 8  & 2   & 13 & 0   & 3 & 2   & 2  & -   & -  & 2.03 & Some impact                                                          & 0.894                                                \\
 \cellcolor[gray]{0.85}                                             & Confidence & 24 & 0   & 5  & 0   & 1 & 0   & 0  & 0   & 0  & 1.23 & Very confident                                                       & 0.496                                                \\
\greyrow                                                            & Workload   & 13 & 2   & 11 & 3   & 1 & -   & -  & -   & -  & 1.62 & Some increase                                                        & 0.601                                                \\
\cellcolor[gray]{0.85} \multirow{-4}{*}{GPWS}                       & Trust      & 10 & 3   & 16 & 0   & 1 & 0   & 0  & 0   & 0 & 1.65 & Some distrust  & 0.519
\\ \bottomrule
\end{tabular}
\end{table*}
\end{document}